\documentclass[twocolumn,showpacs,showkeys,preprintnumbers,amsmath,amssymb]{revtex4}

\usepackage{graphicx}
\usepackage{dcolumn}
\usepackage{bm}

\newcommand{\newc}{\newcommand}
\newc{\ra}{\rightarrow}
\newc{\lra}{\leftrightarrow}
\newc{\beq}{\begin{equation}}
\newc{\eeq}{\end{equation}}
\newc{\barr}{\begin{eqnarray}}
\newc{\earr}{\end{eqnarray}}

\begin{document}
\vspace*{2.0 cm}
\preprint{APS/IOA-165}

\title{Simulated neutrino signals of low and intermediate energy neutrinos on Cd detectors}

\author{ J. Sinatkas}
\email{ sinatkas@kastoria.teiwm.gr}
\affiliation{Department of Informatics Engineering, Technological Institute of Western Macedonia, Kastoria, GR-52100}$^{3}$
\author{ V. Tsakstara}
\email{ vtsaksta@cc.uoi.gr }
\affiliation{Electrical Engineering Department, Technological Institute of Western Macedonia, School of Applied Science,Kozani, GR-50100 }
\affiliation{Division of Theoretical Physics, University of Ioannina, GR-45110 Ioannina, Greece}
\author{Odysseas Kosmas}\email{odysseas.kosmas@manchester.ac.uk}
\affiliation{Modelling and Simulation Centre, MACE, University of Manchester, Sackville Street, Manchester, UK}

\date{\today}

\begin{abstract}
Neutrino-nucleus reactions cross sections, obtained for neutrino energies in the range
$\varepsilon_{\nu}\leq 100-120$ MeV (low- and intermediate-energy range), which refer to promising 
neutrino detection targets of current terrestrial neutrino experiments, are presented and discussed. 
At first, we evaluated original cross sections for elastic scattering of neutrinos produced from 
various astrophysical and laboratory neutrino sources with the most abundant Cd isotopes $^{112}$Cd, 
$^{114}$Cd and $^{116}$Cd. These isotopes constitute the main material of the COBRA detector aiming 
to search for neutrinoless double beta decay events and neutrino-nucleus scattering events at the 
Gran Sasso laboratory (LNGS). The coherent $\nu$-nucleus reaction channel addressed with emphasis 
here, dominates the neutral current $\nu$-nucleus scattering, events of which have only recently 
been observed for a first time in the COHERENT experiment at Oak Ridge. Subsequently, simulated 
$\nu$-signals expected to be recorded at Cd detectors are derived through the application of modern 
simulation techniques and employment of reliable neutrino distributions of astrophysical $\nu$-sources 
(as the solar, supernova and Earth neutrinos), as well as laboratory neutrinos (like the reactor neutrinos, 
the neutrinos produced from pion-muon decay at rest and the $\beta$-beam neutrinos produced from the 
acceleration of radioactive isotopes at storage rings as e.g. at CERN).
\end{abstract}


%
\maketitle
%

\section{Introduction}
In recent interdisciplinary investigations in nuclear, particle and astro-particle 
physics, the interactions of neutrinos with matter play key role in understanding 
deeply the underlying physics. Exact measurements and reliable models of neutrino-matter 
interactions provide unquestionable requirements for unravelling top physics issues as 
neutrino properties, neutrino oscillations, supernova dynamics, dark matter detection
and many others \cite{Ejiri-PR00,Ejiri-plb-02,Zuber-plb-1,Zuber-plb-2}. To enable further 
progress, relevant nuclear model calculations, across a wide energy range and in various nuclear 
isotopes, may provide significant results \cite{DonPe,Kos-Ose,kolbe-kos}. 

Recently, the neutrino-nucleon and neutrino-nucleus cross-section uncertainties have reached
a limiting factor in the judgement of neutrino interaction models and in interpreting many 
neutrino experiments \cite{Ejiri-PR00,Ejiri-plb-02,Zuber-plb-1,Zuber-plb-2} and specifically 
experiments like the COHERENT where recently coherent neutrino nucleus scattering events have 
been measured for a first time \cite{CEnN_neut,Scholberg,Scholb_IoP-2010}. Furthermore, the 
presence of 
important nuclear effects impact the interaction cross sections as well as the final nuclear 
states reached through the scattering process \cite{TSAK-prc1,kos-eric-11,tsak_prc2}. 
The nuclear effects also affect the rebuilding of the incoming $\nu$-energy spectra of the
neutrino sources that are key-role input for the resolution of neutrino detection signals. 
Understanding neutrino-nucleus scattering processes provides to experimentalists good
information to separate the background events from the detection signal 
\cite{KamLAND,KamLAND-11,Borexino-Col,Borexino,Borexino-13,Zub-SNO+,LENA}.

The current neutrino physics searches are categorized according to the incident neutrino 
energy in the scattering process. Thus, the range below about 10 MeV (low-energy, from a 
nuclear physics viewpoint) is connected to Geo-neutrino and solar neutrino studies
\cite{KamLAND,KamLAND-11,Borexino-Col,Borexino,Borexino-13,Zub-SNO+}, 
the neutrino energy range of 10 up to about 100-120 MeV (intermediate energy) covers a 
set of $\nu$-physics topics in the front of nuclear structure physics and 
astro-particle physics such as core-collapse supernovae dynamics and dark matter detection 
\cite{TSAK-prc1,kos-eric-11,tsak_prc2,Kolbe-Lag-Pin-03,Athar-Sing}, while the energy range 
from 0.1-0.2 GeV up to about 10 GeV is related to meson decay neutrino beams such as 
those employed for long-baseline (high energy) neutrino experiments 
\cite{bet-beam-Zuc,bet-beam-Vol,Volpe07,Zuber-PPNP-10}. 

Due to the fact that neutrinos interact very weakly, they are unique messengers from 
astrophysical sources (the Earth, the Sun, the supernovae and other stars) 
\cite{Ejiri-PR00,Ejiri-plb-02,Kolbe-Lag-Pin-03} allowing us to investigate deep into 
the astrophysical objects \cite{Kolbe-Lag-Pin-03,Smpon-Ody-2015,Smpon_Ody-2017,Smpon-Ody-2018}. 
In the near future, 
remarkably sensitive detectors as liquid-scintillator detectors, liquid argon time 
projection chambers and water-Cherenkov detectors would operate aiming to study neutrino 
physics issues of astrophysical neutrino sources \cite{Zub-SNO+,LENA,Zuber-PPNP-10} (for 
higher energy neutrinos, like e.g. those coming from active galactic nuclei, black hole 
binary stars, etc. operating detectors as IceCube, KM3Net and others are appropriate) 
\cite{IceCube_2015,KM3Net_2016}. Each detector type has specific advantages (e.g. for 
supernova neutrinos, a combination of all types may allow for a better investigation of 
the relevant open issues).

Our present work focuses on the interpretation of various $\nu$ signals generated in nuclear 
detectors of terrestrial neutrino experiments through the investigation of the nuclear response 
of Cd detector materials to various neutrino energy spectra 
\cite{Zuber-plb-1,Zuber-plb-2,Zuber-PPNP-10,Tsak_AHEP}. We emphasize on signals coming from 
geo-neutrinos, solar-neutrinos, supernova-neutrinos, reactor-neutrinos and neutrinos 
generated from the decay of stopped pions and muons. 

The main ingredients to this aim are: 
(i) The original differential and integrated cross sections of the neutral-current 
reactions of neutrinos, $^{112,114,116}$Cd$(\nu, \nu')^{112,114,116}$Cd$^*$, and anti-neutrinos, $^{112,114,116}$Cd$(\widetilde{\nu}, \widetilde{\nu}')^{112,114,116}$Cd$^*$, computed for the 
coherent channel by using a refinement of the quasi particle random phase approximation (QRPA)
\cite{Tsak_AHEP,TSAK-prc1,tsak_prc2,Kosmas-94,Kosmas01}.
(ii) Reliable descriptions of the shapes of neutrino energy distributions coming out of 
numerical simulations of distributions in neutrino-energies $\varepsilon_{\nu}\leq 100-120$ MeV
(for the above mentioned $\nu$-sources). 
(iii) Modern computational tools \cite{ody05,ody01,ody02,ody03,ody04} for the required folding (convolution) procedure in order 
to simulate the 
signal expected to be recorded on the Cd detectors CdTe or CdZnTe (the detector media of 
COBRA experiment) \cite{Zuber-plb-1,Zuber-plb-2,Zuber-PPNP-10} from neutrino sources as the
geo-, reactor-, solar-, supernova- and pion/muon decay neutrinos.
We mention that, the response of the Cd isotopes in the particle-bound excitation region, 
which coincides with the energy range of geo-neutrinos, is rather rich and this is one of 
the motivations for performing the present calculations. The next generation detectors 
(LENA, Borexino, SNO+) \cite{Zub-SNO+,LENA}, are expected to give useful answers to 
several questions of geological importance regarding the precise geo-$\nu$ fluxes 
and abundances of natural radioactive elements ($K$, $U$, $Th$) in the Earth's interior
\cite{Vogel-Beacom,Dye-2011,Fiorentini-2003,Fiorentini-2010}.

In this work we pay special attention on the coherent elastic neutrino-nucleus scattering 
(CEvNS) that is a process in which the target nucleus recoils coherently via a combined 
neutral current exchange width with neutrinos or anti-neutrinos. This process is well 
predicted by the standard model of the electroweak interactions and has large cross sections 
(10$^{-39} \, {\textrm cm}^2$ in the neutrino-energy region ($\varepsilon_\nu \le$50 MeV). 
This process has very recently been observed in the COHERENT experiment at a 6.7 $\sigma$ 
confidence level (CL), by using a low-background CsI[Na] scintillator 
\cite{CEnN_neut,Scholberg,Scholb_IoP-2010}. The detector was exposed to a $\nu_{\mu}$ 
neutrino beam coming from the Spallation Neutron Source 
(SNS) at Oak Ridge, USA \cite{Scholberg}. This facility generates the most intense 
(pulsed) neutron beam in the world while simultaneously a significant yield of neutrinos 
is generated when pions (product of proton interactions in the target) decay at rest
(prompt neutrinos). In addition, the muons produced from the charged-pion decay generate 
the known as delayed neutrino beam \cite{Scholb_IoP-2010}.   

Even though many groups world-wide are now studying the difficult low-energy nuclear recoil 
signature, only a few sources, in specific nuclear reactors and spallation neutron sources 
yield the required neutrino-energy beams in adequate quantities for such measurements
\cite{ORLaND,Avignone03,Burman-03}. In our present theoretical work, we do not address the 
improved constraints derived from this dataset on non-standard neutrino interactions with 
quarks (for a comprehensive discussion on this issue the reader is referred e.g. to Refs. 
\cite{Papoul-Kosm,Papoul_tsk_AHEP-2015} and references therein). The present article is an 
extension of our previous calculations performed in Ref. \cite{TSAK-prc1,tsak_prc2,Tsak_AHEP} 
and we used the same but slightly improved nuclear method. The extension refers to the 
employment of new detector isotopes and the better accuracy of the calculations 
\cite{TSAK-prc1,kos-eric-11,tsak_prc2,Tsak_AHEP}. 

In the rest of the paper, at first (Sections 2 and 3), the main formalism is described and 
original cross sections calculations are presented. Then (Sections 4 and 5), a description of 
the main characteristics of the low and intermediate energy neutrino sources addressed here 
are briefly summarized and folded cross sections as well as event rates for neutral current 
neutrino scattering off the $^{112}$Cd, $^{114}$Cd and $^{116}$Cd isotopes are presented and 
discussed. Finally (Section 6), the main conclusions of the present work are extracted.

\section{Brief Description of the Formalism}

\subsection{ Angle differential coherent $\nu$-nucleus cross section }

In the description of the $\nu$-nucleus scattering, the angle differential cross section $d\sigma/d\Omega$
is a useful quantity. For the elastic-scattering of a neutrino with energy $\varepsilon_\nu$ on a nucleus 
(A,Z) the angle differential cross section (with respect to the the scattering angle $\vartheta$) is defined 
as \cite{Engel-91,De-Vries,Drukier,Papoul-Kosm}
\begin{equation}
\frac{d\sigma}{d\Omega}=\frac{{\mathrm G}^2_{\mathrm F}}{4\pi^2} \varepsilon^2_\nu (1 + \cos\vartheta) 
\frac{Q^2_w}{4} {\cal F}(q^2)^2 \,   \label{dsigma_dOmega}
\end{equation}
($G_F =1.1664\times 10^{-5}$ GeV$^{-2}$ is the Fermi weak coupling constant). In this definition, 
the quantity ${\cal F}(q^2)$ includes the nuclear structure dependence of the cross section as
\cite{Kosmas01,Kosmas-94}
\begin{equation}
{\cal F}(q^2) = \frac{1}{Q_w}\Bigg[(1-4 \sin^2\Theta_w) Z F_Z(q^2) - N F_N(q^2) \Bigg] \,  
\label{fQQ}
\end{equation}
where ${\Theta_w}$ denotes the weak mixing angle, known as Weinberg angle which takes the value 
sin$^2\Theta_w \approx$ 0.2313.
In Eqs. (\ref{dsigma_dOmega}) and (\ref{fQQ}), $Q_w$ denotes the weak charge of the target nucleus
given by
\begin{equation}
Q_w = (1-4\sin^2\Theta_w)Z - N \, .
 \label{Qw}
\end{equation}

The latter expression shows that, the neutron coherence of neutral currents (NC), in the case of neutron 
rich targets, provides large cross sections. 
This effect can be exploited in detecting, e.g. earth and sky neutrinos by measuring nuclear recoils. 
Measurements of these (NC) cross sections may also provide useful information about the neutrino source 
\cite{Kolbe-Lag-Pin-03} and yield information about the primary neutrino fluxes, i.e. before flavour 
conversions in the neutrino sphere of core collapse supernovae. 

The sensitivity of the coherent scattering channel to the neutron number in the target nucleus, may 
provide nuclear structure information through investigation of $\nu$-nucleus scattering and the 
possibility to search for non-standard neutrino physics by taking advantage of the flavour-blind 
nature of the process \cite{Papoul_tsk_AHEP-2015,Horowitz-PRC-2012}. 

The ground-state elastic nuclear form factors, $F_{Z}(q^2)$ for protons and $F_{N}(q^2)$ for neutrons
entering Eq. (\ref{fQQ}), are defined by
\begin{equation}
F_k(q^2) = \frac{k}{4\pi} \int {j_0(q r)\rho_{n,p}(r)r^2 dr} \, , \, \, k = {N,Z} \label{fnQQ}
\end{equation}
and are normalized as $F_{N,Z}(q^2= 0) = 1$. In the latter equation, $\rho_{n,p}(r)$ denote the 
neutron (n) and proton (p) charge density distributions 
with $j_0(qr) = {\sin(qr)}/{(qr)}$ being the zero-order spherical Bessel function (we neglect a 
small correction from the
single-nucleon form factors proportional to $e^{-(q b_N)^2/6}$ with $b_N \approx 0.8$ fm being 
the nucleon harmonic oscillator size parameter \cite{tsk-NPA-92}). The proton density $\rho_{p}(r)$ 
is often taken from experiment whenever measured charge densities are available \cite{Kosmas-94,Engel-91}. 

Moreover, assuming that $F_N\approx F_Z$, from Eqs. (\ref{dsigma_dOmega}) and (\ref{fQQ})
(in nuclei with $J^\pi = 0^+$ ground state), one obtains 
\begin{equation}
\frac{d\sigma (\varepsilon_{\nu},\vartheta)}{dcos\vartheta} = \frac{G^2_F}{2\pi} 
(1 + \cos\vartheta) \varepsilon^2_\nu \Bigg[ f_V^p Z + f_V^n N \Bigg]^2 F_Z^2 (q^2) \, . 
\label{dsigma_dtheta}
\end{equation}
where $f_V^p$ and $f_V^n$ stand for the polar-vector couplings of the weak neutral current 
\begin{equation}
f_V^p  = \frac{1}{2} - 2 \sin^2\Theta_W   \, , \qquad  f_V^n  = -\frac{1}{2} \, .
\label{Vect-Couplin}
\end{equation}
Thus, the coherent cross section depends on the square of the ground-state nuclear form factor 
${\cal F}(q^2)$ at momentum transfer $q$ given by
\begin{equation}
q = 2\varepsilon_\nu \sin(\vartheta/2) \, ,
 \label{QQ}
\end{equation}

From Eq. (\ref{fQQ}), we see that, since $f_V^p  = (1-4\sin^2\Theta_w) / 2 \approx 0.0374$ is small, 
a neutrino scattered elastically on a spin-zero nucleus couples mostly to the neutron distribution, 
$\rho_n(r)$. A measurement of the cross section for this process would, at some level, provide a
determination of the neutron form factor $F_N(q^2)$ \citep{Horowitz-PRC-2012,Chiang-Oset-Kosmas}.
Some authors consider that this would be complementary to parity violating experiments 
\citep{Horowitz-PRC-2012,De-Vries} because it would provide 
additional data, obtained at different energy ranges and with different nuclei 
that could be used to calibrate nuclear structure calculations 
\cite{Kosmas01,Kosmas-94,Engel-91,De-Vries,Drukier}.

In earlier astrophysical estimations of the coherent scattering cross sections within the 
Standard Model (SM) \cite{Drukier,Bachall} (also in recent beyond the SM calculations
\cite{Omar-1,Omar-2}), the approximation $F_N (q^2) \approx F_Z (q^2)\approx 1$ was used for 
the total coherent cross section $\sigma_{tot}(\varepsilon_{\nu})$ written as
\begin{equation}
\sigma_{tot}(\varepsilon_{\nu}) =  \frac{G^2_F}{8\pi}\Bigg[ (1 - 4\sin^2\Theta_W)Z - N \Bigg]^2 
\varepsilon^2_\nu \, .
 \label{dsigma_tot}
\end{equation}
(we mention that available experimental data for neutron form factors  are very limited).


From an experimental point of view, and particularly for the neutrino facilities near spallation 
sources \cite{Avignone03,JDV-Avig}, it is also interesting the expression of the coherent 
differential cross section as a function of the nuclear recoil energy $T_A$. This is 
approximately written as \cite{Giom-JDV,JDV-Giomat,JDV-Avig,Vogel-Engel}
\begin{equation}
\frac{d\sigma (\varepsilon_{\nu}, T_A)}{dT_A} = \frac{G^2_F}{4\pi}Q_W^2M 
\left( 1 - \frac{M T_A}{2\varepsilon^2_\nu} \right)F(2MT_A^2) \, ,
 \label{dsigma-dT_A}
\end{equation}
where M is the nuclear mass and F denotes the ground state elastic form factor of the 
target nucleus. For the sake of completeness, we note that other expressions, 
including higher order terms with respect to $T_A$ can be found, see e.g. Refs. 
\cite{Scholberg,Scholb_IoP-2010,Giom-JDV,JDV-Giomat,JDV-Avig}. The contribution, however, 
of these therms is negligible and thus, higher order terms in Eq. (\ref{dsigma-dT_A}) does 
not influence essentially the calculations. Our present coherent differential cross sections 
are not obtained as functions of the recoil energy but as functions of the scattering angle 
or the momentum transfer connected through Eq. (\ref{QQ}). 

It should be noted that, the signal on the coherent neutrino-nucleus scattering experiments 
is significantly different compared to that of the incoherent scattering where the signal 
could be an outgoing particle or a de-excitation product \cite{Tsak_AHEP}.

The total coherent cross section $\sigma_{tot}(\varepsilon_{\nu})$ is obtained by integrating 
numerically Eq. (\ref{dsigma_dtheta}) over the angle $\theta$ ($\theta_{min}=0$ to $\theta_{max}=\pi$) 
or Eq. (\ref{dsigma-dT_A}) over $T_A$ between 
$$
T_A^{min}=\frac{T_A}{2}+\sqrt{\frac{T_A}{2}(M_A 
+ \frac{T_A}{2})} \, ,
$$ 
to $T_A^{max} = \infty$ \cite{Papoul-Kosm,Vogel-Engel,Kolbe-96}. 

Before closing this sections, it is worth mentioning that, in our present calculations of 
the neutrino-nucleus cross sections part of the cross-section uncertainties are removed by 
performing realistic nuclear structure calculations for both proton and neutron nuclear form 
factors (for a recent comprehensive discussion on this issue the reader is referred e.g. to 
Ref. \cite{Papoul_tsk_AHEP-2015} where the results coming out of different nuclear models 
and various approximations are presented and discussed).

\section{Original cross section calculations}

The neutral-current scattering of low and intermediate energy neutrinos $\nu_\ell$ or anti-neutrinos 
$\widetilde{\nu}_\ell$ ($\ell = e, \mu, \tau$) off the $^{112,114,116}$Cd isotopes (with abundances 
24.13\%, 28.8\% and 7.5\%, respectively, the first two are the most abundant Cd isotopes) are represented 
by 
\begin{equation}
 \nu_l({\widetilde{\nu}_l}) + \, ^{112,114,116}Cd \rightarrow \, ^{112,114,116}Cd^* + {{\nu}^\prime_l}({\widetilde{\nu}^\prime_l}) \, ,
\label{neut-Cd}
\end{equation}
(Cd$^*$ denote excited states of Cd-isotopes). We mention that, 
the above reactions of the Cd-isotopes and also the charged-current (CC) reactions for $\ell = e$, 
play significant role in astrophysical 
environment by affecting the electron fraction $Y_e$ of the matter and its strong effect on the matter 
flow \cite{Lang-Pin-03,Athar-Sing,Juoda-Lang,Lang-08,Hax-87}. 

In the first step of the present calculations, we evaluate original cross sections 
for the coherent channel (ground state to ground state transitions) of the 
reactions of Eq. (\ref{neut-Cd}) \cite{DonPe,TSAK-prc1,tsak_prc2,Don-Wal72,Don-Wal73,Kolbe-96}. 
As can be seen from Eq. (\ref{dsigma_dtheta}), the original cross section for scattering of neutrinos, 
$\nu_l$ or anti-neutrinos, $\widetilde{\nu}_l$, are identical (this holds only for the coherent channel).
The signal (folded cross section) on the nuclear detector, however, as we will see in Sections IV and V,
could be significantly different. This is due to the flavour dependent energy distributions of the 
$\nu$-beam reaching the nuclear detector, that enters in the folding procedure.

In this work, the required nuclear ground state wave functions are obtained from mean-field 
calculations using the successful Woods-Saxon interaction plus the monopole (pairing) interaction of 
the Bonn C-D potential. The ground state of the studied (even-even) $^{112,114,116}$Cd 
isotopes (they have ground state spin $|J_i^{\pi_i}\rangle = |0^{+}_{gs}\rangle$) is computed by solving 
iteratively the BCS equations \cite{TSAK-prc1,tsak_prc2,prc3,Papoul-Kosm,Tsak_AHEP}. 

In Table \ref{Table1}, we list the values of the resulting pairing parameters ($g_{pair}^{p,n}$) and 
the (theoretical) energy gaps ($\Delta^{th}_{p,n}$) for protons ($p$) and neutrons ($n$) determined 
at the BCS level for the above isotopes. As is well 
known, these parameters renormalise the pairing interaction of the Bonn C-D potential in order to fit 
the theoretical gaps, $\Delta^{th}_{p,n}$, to the empirical ones $\Delta^{exp}_{p,n}$. The latter 
are provided through the application of the three point formulas (see Appendix) by using the empirical
separation energies (for protons and neutrons, $S_{p,n}$) of the neighbouring nuclear isotopes 
\cite{tsak_prc2,Tsak_AHEP}. The values of the $g_{pair}^{p,n}$ adjust reliably the empirical 
energy gaps (see Table \ref{Table1}) \cite{TSAK-prc1,tsak_prc2,prc3,Bal-Ydr-Kos-11,Tsak_AHEP}.

\begin{table*}
\begin{center}
\begin{tabular}{|c|c|c|c|c|c|c|c|c|c|c|}
\hline\hline
&&&&&&&&&\\
Isotope&Z, N& Abundance ($\%$)& {b (fm) }& $g_{pair}^n$ & $g_{pair}^p$
& $\Delta^{exp}_{p}$ & $\Delta^{th}_{p}$ & $\Delta^{exp}_{n}$ & $\Delta^{th}_{n}$ \\
 \hline
&&&&&&&&&\\
 $^{112}\!Cd$ & 48, 64& 24.13 & 2.208 & 1.001 & 1.064 & 1.516  & 1.512 & 1.320  & 1.322 \\
 $^{114}\!Cd$ & 48, 66& 28.73 & 2.214 & 0.956 & 0.975 & 1.441  & 1.441 & 1.351  & 1.351 \\
 $^{116}\!Cd$ & 48, 68&  7.50 & 2.219 & 1.069 & 1.043 & 1.432  & 1.432 & 1.371  & 1.372 \\
\hline
\hline
\end{tabular}
\caption{ Pairing parameters $g_{pair}^p$ (for protons), and $g_{pair}^n$ (for neutrons) determining 
the monopole pairing interactions for each of the studied isotopes. The obtained theoretical 
values of the energy gaps (in units of MeV), $\Delta^{th}_{p}$ (for protons) and $\Delta^{th}_{n}$ 
(for neutrons), are
also shown for comparison with the empirical ones. As can be seen, the corresponding empirical energy 
gaps, $\Delta^{exp}_{p,n}$ are well reproduced. Values of the harmonic oscillator size parameter, $b$, 
for each of 
the isotopes $^{112,114,116}\!$Cd are also given in this Table.}
\label{Table1}
\end{center}
\end{table*}

The needed proton and neutron nuclear form factors in the context of QRPA are calculated from 
the expressions
\begin{equation}
F_k (q^2) = \frac{1}{k} \sum_{j} {\hat j} \langle (n \ell)j\vert j_0(q r) \vert (n \ell)j \rangle (V_j^k)^2
 \, , k = {N,Z}
\label{BCS-form-fact}
\end{equation}
($V_j^k$ denotes the probability amplitude for proton or neutron occupancies of the single particle 
$(n \ell)j$-level). The summation, runs over the 15 active levels of the chosen model space (the same 
for proton and neutrons) as well as over the fully occupied $j$-levels for which $V_j^k = 1$ (they 
describe a $^{40}$Ca closed core). The model space assumed consists of the major harmonic oscillator 
shells having quantum numbers N=$3, 4, 5$ (N=$2n+\ell$).

\begin{figure}
\includegraphics[scale=1.20]{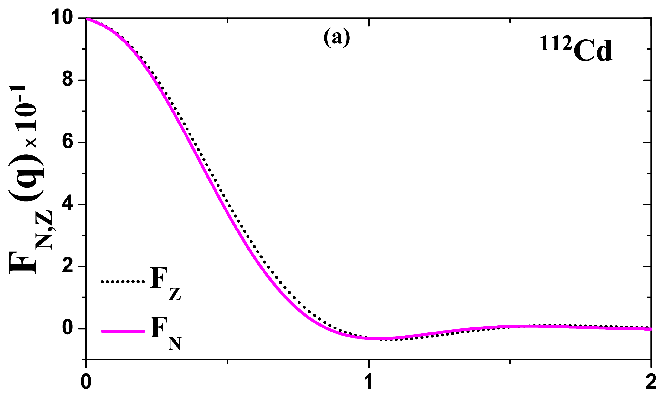}
\includegraphics[scale=1.20]{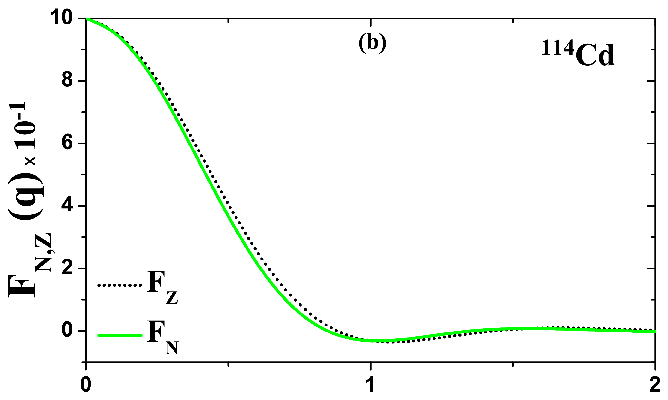}
\includegraphics[scale=1.20]{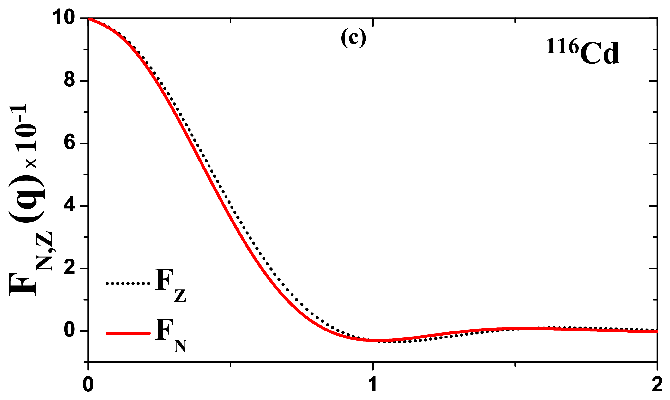}
\includegraphics[scale=1.30]{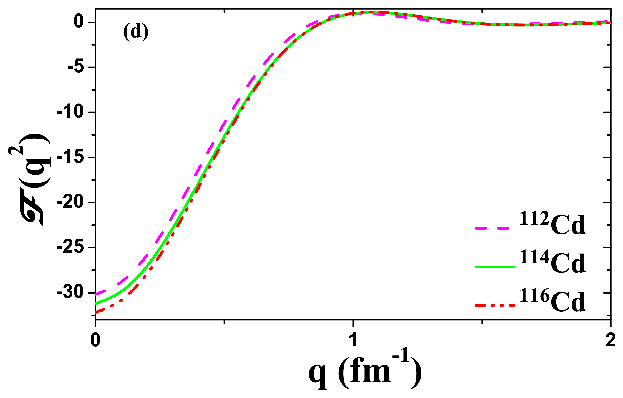}
\caption{ Neutron and proton nuclear form factors $F_{N,Z}(q^2)$ (a) for 
$^{114}$Cd, (b) for $^{112}$Cd, and (c) for $^{116}$Cd isotopes. 
(d) The ground-state elastic nuclear form 
factor ${\cal F}(q^2)$ for $^{112,114,116}$Cd isotopes. } 
\label{Fig1}
\end{figure}
\begin{figure}
\includegraphics[scale=1.30]{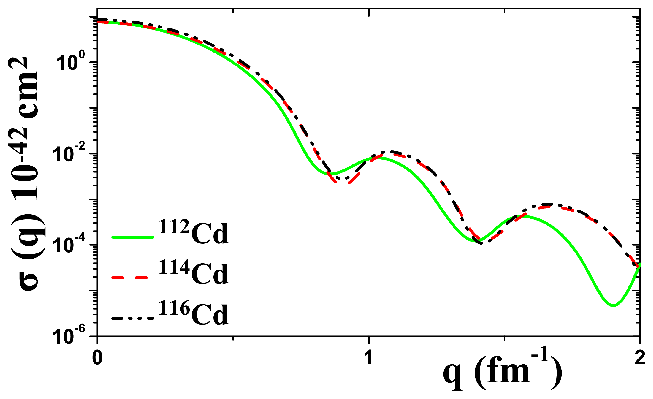}
\includegraphics[scale=0.275]{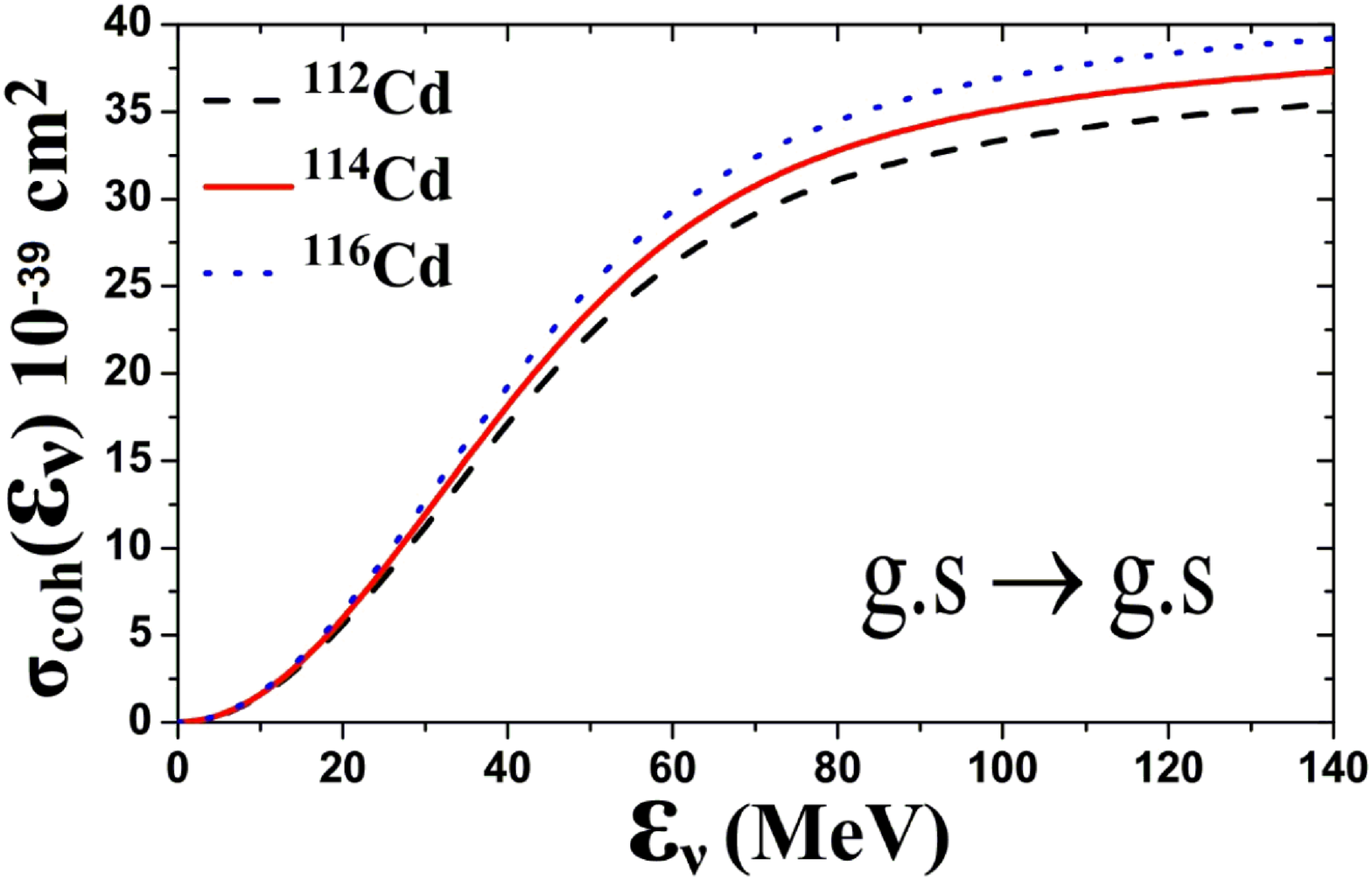}
\caption{ Total cross sections of coherent (ground state to ground state $g.s. \to g.s.$) 
transitions for the neutral current reactions $^{112,114,116}$Cd($\nu_l, \nu_l')^{112,114,116}$Cd$^*$, 
$l=e, \mu, \tau$.}
\label{Fig2}
\end{figure}

In Fig. \ref{Fig1}, the quantities needed for calculating the differential and integrated coherent 
cross section (see Eqs. (\ref{dsigma_dOmega}) and (\ref{dsigma_dtheta})) for the neutrino reactions 
(\ref{neut-Cd}) are illustrated. Figure \ref{Fig1}(a)-(c), shows the form factors for protons $(F_Z)$ 
and neutrons $(F_N)$ obtained with our BCS calculations (for the three isotopes $^{112,114,116}$Cd) 
and Fig. \ref{Fig1}(d) shows the momentum dependence of ${\cal F} (q^2)$ that enters Eqs. 
(\ref{dsigma_dOmega}) and (\ref{dsigma_dtheta}). 

It should be noted that, the corrections due to the nucleon finite size 
($e^{-(qb_N)^2/6}$) and the nuclear center-of-mass motion ($e^{(qb)^2/4 A}$), which enter as an overall 
q-dependent factor in the $F_{N,Z}(q)$, for the medium heavy 
Cd-isotopes are negligible and have been ignored. The correction due to the nucleon finite size 
(the larger of the two) is very well known, but not essential. For small q the 
influence is close to zero while at the maximum momentum q it is about 5\% \cite{tsk-NPA-92}.

As can be concluded from Fig. \ref{Fig1}, the above ground state properties of the three Cd isotopes 
studied are to a large extent similar which means that their nuclear structures are not significantly 
different (all of them have ground state spin $J^{\pi}=0^+$). The differences, are mostly due to 
the small ratio ($\Delta N_i/N \approx 3\% - 6\%$) in their neutron number.

Figure \ref{Fig2} illustrates the total integrated coherent cross sections of $\nu$-$^{112,114,116}$Cd 
scattering as a function of (i) the momentum transfer $q$, Fig. \ref{Fig2}(a), and (ii) the incoming 
neutrino energy $\varepsilon_\nu$, Fig. \ref{Fig2}(b). As mentioned before, these original cross 
sections will be used below for evaluations of flux averaged folded cross sections for various 
neutrino spectra.

Before closing this section, it is worth mentioning that, in calculating the nuclear form factors
${\cal F}(q^2)$, see Fig. \ref{Fig1}, in the context of the QRPA method, the estimated error at 
low momentum transfer is very small, while in the momentum range of our interest 
$0 \le q \le 2 fm^{-1}$, it is at maximum 10-15\%. 
On the other hand, the experimental accuracy, for the proton form factors entering 
Eq. (\ref{fQQ}), usually they come from electron scattering measurements, is of the order of 
1\% \cite{De-Vries}. For neutron form factors, however, the available experimental data are 
limited and, in general, authors discuss about differences between corresponding proton and 
neutron nuclear form factors (in medium heavy isotopes like $^{112,114,115}$Cd) of the order 
of 4 to 8\% \cite{Horowitz-PRC-2012,Chiang-Oset-Kosmas}.

In the next section, we summarize the main features of the $\nu$-energy distributions 
employed in this work for obtaining folded neutrino-nucleus cross sections for each 
$\nu$-source.

\section{Energy-spectra of low-energy and Intermediate $\nu$-sources}

In this section, we focus on the basic characteristics of the currently interesting 
astrophysical (solar-, supernova-, geo-neutrino) and laboratory (reactor neutrino 
and pion/muon decay at rest neutrino) sources, their energy 
spectra of which will be used in the convolution procedure (see next 
section) to obtain convoluted cross sections based on our original cross sections.

In general, the $\nu$-beams of the above mentioned neutrino sources have broad 
energy distributions (sometimes they consist of a mixture of neutrinos and anti-neutrinos) 
characteristic of the considered source. Some well known mono-energetic (monochromatic) 
fluxes are e.g. the one coming out of the charged-pion decay at rest (corresponding to 
the energy $\varepsilon_{\nu_{\mu}}=29.65$ MeV, see Fig. \ref{Spectrumms}(d panel). For the 
non-mono-energetic neutrino fluxes we define the energy distributions $\eta (\varepsilon_\nu)$ 
as
\begin{equation}
\frac{d N_\nu(\varepsilon_\nu)}{d\varepsilon_\nu} \equiv \eta (\varepsilon_\nu)
\end{equation}
($N_\nu$ represents the number of neutrinos of the beam). 

We note that, via these energy spectra $\eta (\varepsilon_\nu)$ of the specific neutrino 
sources, the original $\nu$-nucleus cross sections (of neutral-current reactions) computed 
with the QRPA method, can be connected with physical observables and signals recorded at 
the nuclear detectors through the use of the folding (convolution) method described below. 
The obtained this way folded (convoluted) cross sections represent the simulated nuclear 
detector response of the $^{112,114,116}$Cd isotopes, in the energy range of $\nu$-energy 
distribution of the studied neutrino source. 

The main properties of the aforementioned astrophysical and laboratory neutrinos are 
summarised in the next subsections.

\subsection{Geoneutrinos}

As it is well known, the decay of some radioactive isotopes (mainly $U$, $Th$, $K$) in the 
interior of our planet, makes the Earth a powerful source of low-energy neutrinos in the range 
$\varepsilon_\nu \preceq$ 10 MeV \cite{Vogel-Beacom,Dye-2011,Fiorentini-2003,Fiorentini-2010}. 
Accurate measurements of the flux of these neutrinos 
\cite{KamLAND-11,Borexino-13} are utilized to determine the amount of heat-producing elements 
in the Earth's mantle. This amount may be compared to that estimated through indirect methods,
an information which is important to understand the heat transfer within the Earth. The latter
is responsible for earthquakes and volcanoes. The most recent measurements from 
KamLAND and Borexino \cite{KamLAND, Borexino-Col,Borexino}
are useful to put limits on the parameters of various models describing the structure and
evolution of our planet. 

The Earth neutrinos (mainly electron anti-neutrinos $\widetilde{\nu}_e$), are generated through
$\beta$-decay processes of neutron-rich nuclei like $U$, $Th$ and others. These thermonuclear 
reactions are accompanied by the emission of electrons ($e^-$) and release of energy $Q_\beta$ 
as \cite{Vogel-Beacom}
\begin{equation}
(A,Z)\rightarrow (A,Z+1) + e^- + \widetilde{\nu}_e + Q_\beta\, .
\label{beta-decay}
\end{equation}
A and Z denote the mass and atomic (proton) number, respectively, of the initial (parent) nucleus. 
Part of the decay energy $Q_\beta$ is carried away by anti-neutrinos ($Q_\nu$) while the remainder 
is available for heating ($Q_h$). Thus, $Q_\beta= Q_\nu + Q_h$. 

In general,
the radioactive isotopes of the Earth are classified into three groups: (i) isotopes in the 
$^{238}U$ decay series, (ii) isotopes in $^{232}Th$ decay series, and (iii) $^{40}K$ isotope 
\cite{Vogel-Beacom,Dye-2011}. 
Thus, these isotopes are geologically important because they heat (radiogenic heat) the 
Earth's interior (finally each of them reaches a stable nuclear isotope) via
$\beta$-decays of all intermediate radioactive isotopes.

Figure \ref{Spectrumms}(a panel), shows the individual anti-neutrino spectra from 
$^{40}K$, $^{238}U$ series and $^{232}Th$ series ($\tau_{1/2} = 4.47 \times 10^9$ y, 
$\tau_{1/2} = 14.0 \times 10^9$ y and $\tau_{1/2} = 1.28 \times 10^9$ y, respectively). 
Essentially, these anti-neutrino ($\widetilde{\nu}_e$) energy spectra
come from 82 beta decays in the U series and 70 beta decays in the $Th$ series
\cite{Vogel-Beacom,Dye-2011,Fiorentini-2003,Fiorentini-2010,Anderson-2012,Gouvea-neutrinos}.

\subsection{Solar neutrinos}

The solar neutrino spectra (mainly $\nu_e$ neutrinos) are produced through thermonuclear 
reactions taking place in the interior of the Sun \cite{Bachall-Holstein,Bachall-Urlich,Bachall-05a}. 
The shape of the energy distribution (0.1 MeV$\leq\varepsilon_\nu \leq 18$ MeV) depends on 
the densities and temperatures in 
the Sun's environment \cite{Bachall-Urlich} and the individual process of the reaction chain
(p-p neutrinos, $^{7}$Be neutrinos, $^{8}B$ neutrinos, hep neutrinos, CNO-cycle neutrinos, etc.). 
In Fig.\ref{Spectrumms}(c panel), we show the energy spectra of 
the important $^{8}B$ \cite{Bachall-Holstein} and $hep$ \cite{Bachall-Urlich,Bachall}
neutrino sources predicted by the standard solar model \cite{Bachall}.
The $^{8}B$ $\nu$-spectrum, is nearly symmetric, with a peak at 6.4 MeV while the 
$hep$ spectrum is peaked at 9.6 MeV \cite{Bachall}.

The detection of the solar neutrinos (produced either via the pp-chain reactions or via the 
CNO-cycle processes) by terrestrial experiments (SNO+ \cite{Zub-SNO+,Borexino}), constitutes 
excellent probes for astrophysics, nuclear physics, and particle physics searches
\cite{Bachall-Urlich,Bachall-05a}. Besides the huge success of the solar-neutrino experiments
the last decades, there are still many unsolved questions related to the metallicity 
of the Sun's core, the total luminosity in neutrinos, the neutrino oscillations, etc. 
\cite{KamLAND,Borexino-Col,Borexino,Zub-SNO+,LENA}.

\subsection{Pion-muon decay at rest neutrino energy distributions}

In muon factories (at J-Park, Fermilab, PSI, etc.), from pion and muon decay at rest (DAR), 
in addition to the monochromatic $\nu$-beam peaked at $\varepsilon_{\nu_\mu}$= 29.65 MeV), 
$\widetilde{\nu}_\mu$ and $\nu_e$ beams (with energy of a few tens of MeV) are created. Such 
intermediate energy neutrino sources, are also the currently available at 
high-intensity proton sources, like the SNS at Oak Ridge, the neutrino beam-line produced 
at Fermilab Booster, the future Project-X facilities at Fermilab, etc.
\cite{KamLAND,Borexino-Col,Borexino,Zub-SNO+,LENA}.  

In the farther future, such high-intensity muon beams would offer a possible site for 
neutrino experiments related to supernova neutrinos and for neutrino-nucleus cross section 
measurements in a great number of nuclei \cite{ORLaND,Avignone03,Burman-03,Volpe07}. 
In the operating pion-muon decay at rest neutrino sources (in Fermilab, at USA, J-PARC, 
at Japan, PSI in Switzerland, etc.) and in the neutrino facilities at the Neutron Spallation 
Source (Oak Ridge, USA), $\nu_e$ neutrinos, and ${\tilde \nu}_\mu$ anti-neutrinos are produced 
from the decay of muons according to the reaction
\begin{equation}
\mu^+ \to e^+ + \nu_e +{\tilde \nu}_\mu \, .   
\label{mu-dec}
\end{equation}
The decaying muons result from the decay of pions at rest ($ \pi^+ \to \mu^+ + {\nu}_\mu $). 
Thus, these neutrino beams are not completely pure as, for example, the $\beta$-beam neutrinos
\cite{bet-beam-Zuc,Volpe07}. The energy-spectra of $\nu_e$ and ${\tilde \nu}_\mu$ neutrinos are 
fitted with the normalized distributions \cite{BOONE,Anderson-2012}
\begin{equation}
\eta_{\nu_e}(\varepsilon_{\nu}) = 96 \varepsilon^2_{\nu} \, M^{\, -4}_\mu\, \left( M_\mu -2 
\varepsilon_{\nu} \right) \, ,
\label{Lab-dist-nnu}
\end{equation}
\begin{equation}
\eta_{{\tilde \nu}_\mu}(\varepsilon_{\nu}) = 16 \varepsilon^2_{\nu} \, M^{\, -4}_\mu \, 
\left( 3 M_\mu - 4\varepsilon_{\nu} \right) \, , \label{Lab-dist-nmu2}
\end{equation}
see Fig. \ref{Spectrumms}(d panel), where $M_\mu=105.6$ MeV, is the muon rest mass. 
The ${{\tilde \nu}_\mu}$ spectrum is peaked at $\varepsilon_{\nu}^{max} = 52.8 $ MeV $ = M_\mu/2$ 
while that of ${\nu_e}$ is peaked at $\varepsilon_{\nu}^{max} = 35.2 $ MeV $ = M_\mu/3$   
\cite{BOONE,kolbe-kos}. 

Obviously, the analytic expressions of Eqs. (\ref{Lab-dist-nnu}) and (\ref{Lab-dist-nmu2}), 
are convenient for the required integrations in the folding procedure, see below 
\cite{TSAK-prc1,tsak_prc2,prc3,Tsak_AHEP}. On the other
hand, their energy range and shape roughly resembles that of SN neutrinos.

\subsection{Reactor Neutrino spectra}

The fission of very heavy nuclear isotopes $^{235}U$, $^{239}Pu$, and $^{238}U$ in the 
nuclear reactors produces a great number of neutron rich nuclear isotopes. Because these 
products are unstable, they decay via $\beta$-decay emitting anti-neutrinos ($\widetilde{\nu}_e$)
\cite{Davis-Vogel-79,Bugey_experiment}. Hence, nuclear reactors, operate as intense 
$\widetilde{\nu}_e$ sources for many experiments, giving fluxes of the order of 
$\sim 10^{13} \, \widetilde{\nu}/$cm$^{2}$ s, at distances $\sim$ 10 m from the reactor core.

\begin{figure}
\includegraphics[scale=1.26]{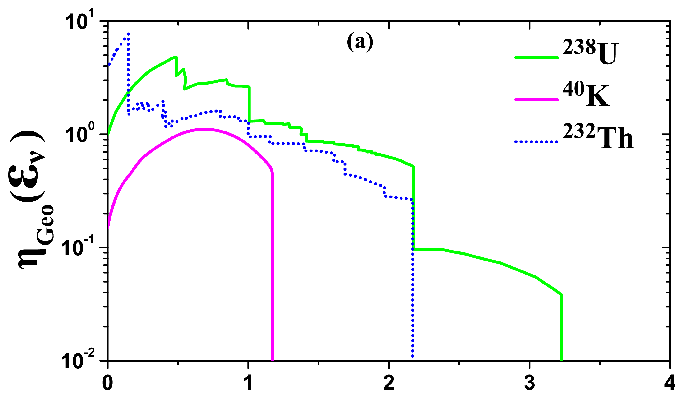}
\includegraphics[scale=1.22]{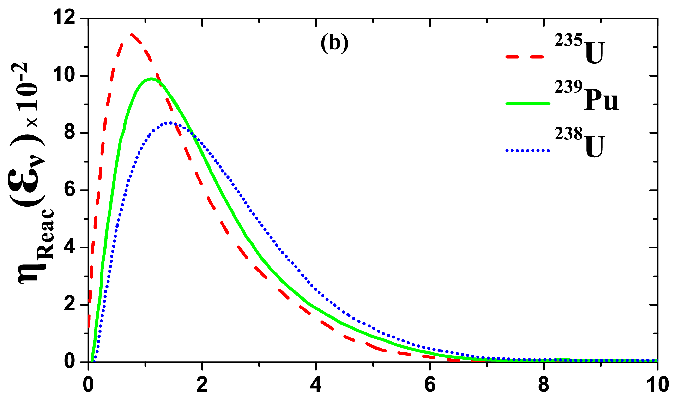}
\includegraphics[scale=1.22]{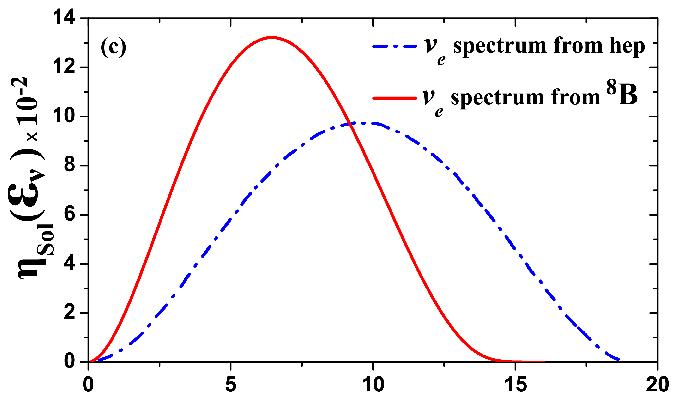}
\includegraphics[scale=1.31]{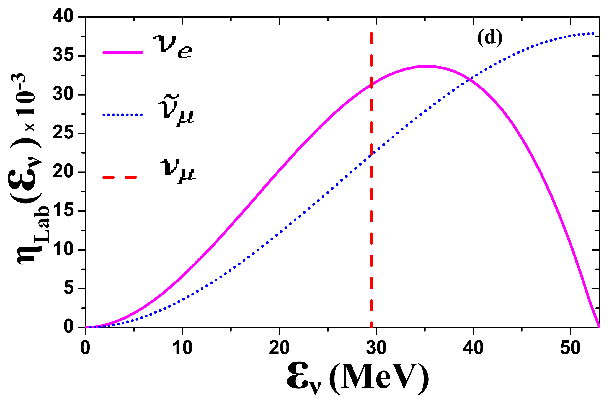}
\caption{(a) Spectra of the U-Series, Th-Series and $^{40}K$ Geo-Neutrinos. Neutrinos from 
$^{40}K$ electron capture are also shown in this figure. (b) Normalized reactor neutrino 
spectra. (c) Normalized energy spectrum of $^{8}B$ and hep $\nu_e$ solar neutrinos. 
(d) Energy-spectra of $\nu_e$ and ${\tilde \nu}_\mu$ neutrino beams, generated from the 
muon-decay at rest (see e.g. Refs. \cite{Anderson-2012,BOONE}).}
\label{Spectrumms}
\end{figure}

The energy spectrum of these anti-neutrinos, characteristic of the $\beta^{-}$ decay spectrum, 
is peaked at very low energies $\sim$ 0.3 MeV and covers the energy region below $\sim$ 
10 MeV. Figure \ref{Spectrumms}(b panel) illustrates the reactor neutrino spectra normalized 
so as the sum over all data-points to be equal to unity. The adopted fuel composition is 
62$\%$ $^{235}U$, 30$\%$ $^{239}Pu$ and 8$\%$ $^{238}U$ \cite{Davis-Vogel-79,Tengblad-89}.

Currently operating reactor neutrino experiments, like the TEXONO experiment in Taiwan
\cite{TEXONO-Kerman,TEXONO-Sevda}, the MINER experiment at the Nuclear Science Center, 
Texas A\&M University (using neutrinos from the TRIGA reactor) \cite{MINER-Agnolet}, are 
excellent probes of beyond the standard model neutrino physics searches (electromagnetic 
$\nu$-properties) and coherent $\nu$-nucleus scattering studies. 

\subsection{Supernova neutrino spectra}

Supernovae (SN) play key role in the development of our Universe, indicated e.g from the fact 
that modern simulations of galaxies formation cannot reproduce the structure of the galactic 
disk without considering supernova data. Today, though the physics of core-collapse supernovae 
is not yet well-understood, investigations of SN neutrinos supply rich information for 
understanding their dynamics, the mechanism of SN-neutrino emission, etc., and for 
interpreting the supernova neutrino burst measurements \cite{DonPe,Kos-Ose,Ejiri-PR00}. 
Multiple physics signatures are expected from a core-collapse explosion in the next supernova 
observation \cite{Burman-03,Kolbe-Lag-Pin-03,Fiorentini-2003,Fiorentini-2010}. The detection 
of a future galactic supernova will provide invaluable information on the astrophysics 
of core-collapse explosion while the high statistics of a galactic SN neutrino signal may 
allow us to unravel the relevant scenarios.

In general, the shape of SN-neutrino energy-distributions is determined by the conditions 
pertaining during their emission from the collapsing star causing the cooling of the 
proto-neutron star formed in its center \cite{Lang-Pin-03,Bethe90,Lank-prl,Janka95,Janka07}.
For the energy distribution of SN neutrinos, some authors used available terrestrial neutrino 
sources with similar energy spectra, like the Neutron Spallation Source neutrinos and the 
boosted radioactive neutrino beams (beta beam neutrinos), in order to test the response of 
some $\nu$-detectors to SN neutrinos \cite{ORLaND,Avignone03,Burman-03}. Recent stellar 
modelling use analytic expressions that include various effects through a chemical potential 
parameter in the well-known two-parameter Fermi-Dirac (FD) distribution \cite{Janka89} 
or through the average $\nu$-energy in the analytically simpler two parameter Power-Law 
(PL) distribution (see Appendix) \cite{Janka89,Raffelt03,Raffelt-proc03}.

Both parametrizations, FD and PL, yield similar distributions characterized by the 
temperature $T$ or the average $\nu$-energy $\langle \varepsilon_{\nu}\rangle$ 
\cite{tsak_prc2,prc3,Duan08,Jachowicz06,Vtsak-Zn-fold,TSAK10-erice2-2010}.
These analytic normalised expressions contain two parameters to include modulation effects 
due to various corrections required to modify the purely thermal shape initially employed \cite{Janka89,Raffelt03,Jachowicz06}. The two parameter FD distribution includes 
the known pinching effect through the degeneracy parameter (the chemical potential divided 
by the neutrino temperature T), $n_{dg}= \mu/T$ which makes the spectrum more narrow 
compared to the purely thermal shape of temperature $T$ (in MeV) \cite{tsak_prc2}. The two 
parameter PL distribution of SN-$\nu$ energy spectrum \cite{Raffelt03,Raffelt-proc03},
contains as parameters the mean neutrino energy $\langle \varepsilon_{\nu}\rangle$ and the parameter 
$\alpha$ which adjusts the width $w$ of the distribution \cite{tsak_prc2,Janka89,Raffelt03,Jachowicz06}
(see Appendix). 

\begin{table*}
\begin{center}
\begin{tabular}{|c|c|c|c|c|c|c|c|}
\hline
\hline    
\multicolumn{8}{|c|}{ }          \\
\multicolumn{8}{|c|}{ Equivalent Fermi-Dirac and Power-Law Supernova Neutrino Spectra } \\
\hline
\multicolumn{3}{|c|}{  }  &  \multicolumn{5}{|c|}{ }          \\
\multicolumn{3}{|c|}{ Parameter }  &  \multicolumn{5}{|c|}{ Temperature  (in MeV) }    \\
\hline    
   &     &       &         &      &      &      &          \\
Width ($w$)  & Pinching ($\alpha$) & Degeneracy ($n_{dg}$) & 
$\langle \varepsilon_{\nu}\rangle =$10 &
$\langle \varepsilon_{\nu}\rangle =$12 & $\langle \varepsilon_{\nu}\rangle =$16 &
$\langle \varepsilon_{\nu}\rangle =$20 & $\langle \varepsilon_{\nu}\rangle =$24  (MeV) \\
\hline
     &     &       &         &      &      &      &          \\
 0.7 & 5.1 &  4.4  &    2.14 & 2.57 & 3.42 & 4.28 & 5.13     \\
 0.8 & 3.7 &  2.7  &    2.58 & 3.10 & 4.14 & 5.17 & 6.20     \\
 0.9 & 2.7 &  1.1  &    2.98 & 3.57 & 4.77 & 5.96 & 7.15     \\ 
\hline
\hline
\end{tabular}
\end{center}
\caption{Corresponding values of parameters for equivalent Fermi-Dirac (FD) and Power-Law 
(PL) distributions (SN neutrino energy spectra) of Fig. \ref{fd-pl}. The selected flavour 
dependent mean $\nu$-energy values (describing the PL distribution), 
$\langle \varepsilon_{\nu}\rangle$ in MeV, have been chosen 
as model values for $\nu_e$ (10-12) MeV, $\widetilde{\nu}_e$ (15-18) MeV and $\nu_x$, where 
x = $\nu_\mu$, $\nu_\tau$, $\widetilde{\nu}_\mu$, $\widetilde{\nu}_\tau$ (22-26) MeV. }
\label{Equiv-Param-FD_PL} 
\end{table*}

\begin{figure}
\includegraphics[scale=0.7]{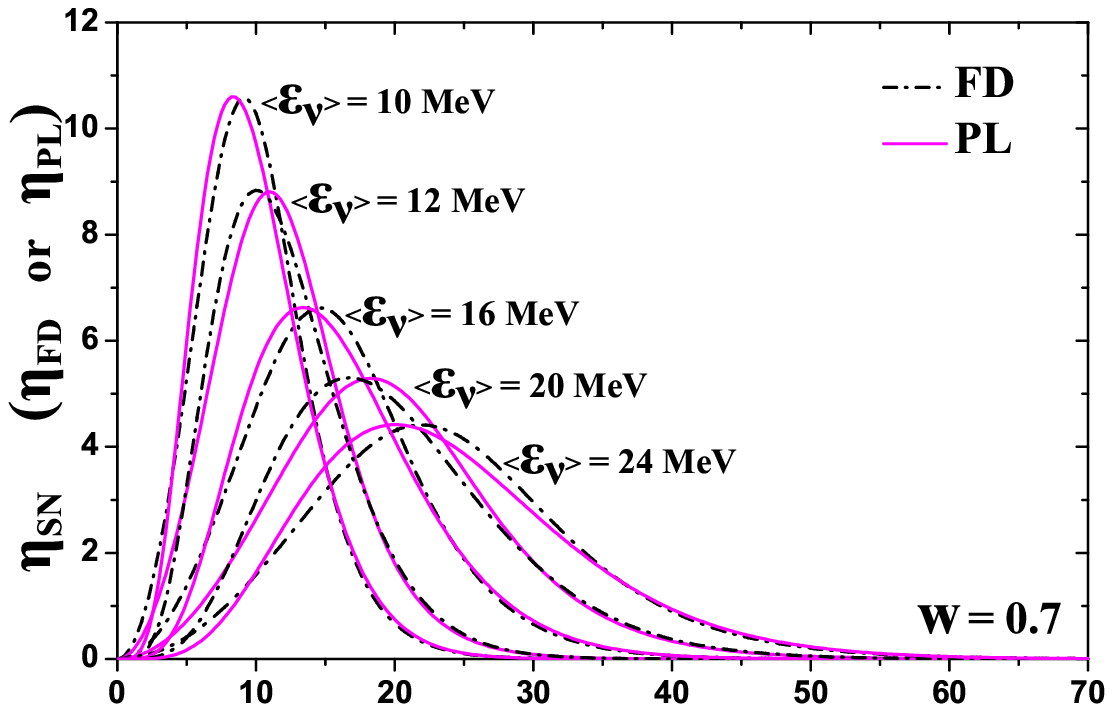}
\includegraphics[scale=0.7]{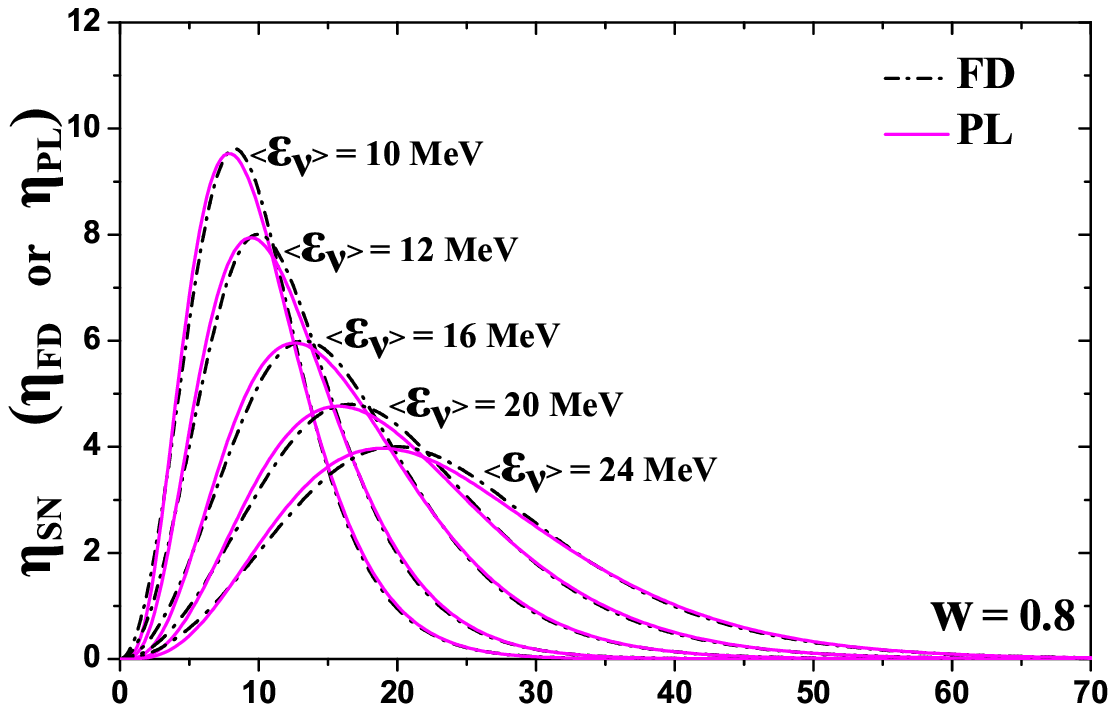}
\includegraphics[scale=0.7]{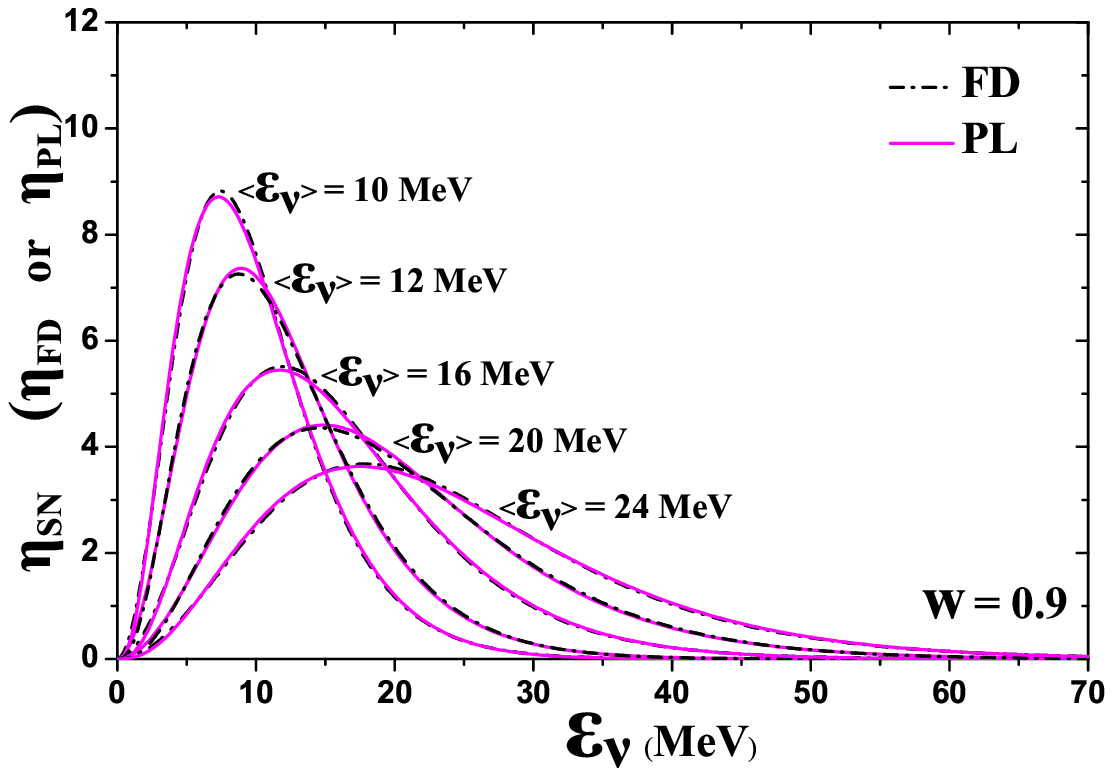}
\caption{ Supernova neutrino energy spectra ($\eta_{SN}$)
coming out of the analytic expressions of:
 (i) the two-parameter Fermi-Dirac distribution (FD) and 
(ii) the two-parameter Power-law (PL) distribution 
(see Appendix). The five sets of values of their 
parameters refer to equivalent distributions (for details see the text). }
\label{fd-pl}
\end{figure}

In Fig. \ref{fd-pl}, some flavour dependent $\nu$-energy spectra ($\eta_{SN}$) 
emitted by a core-collapse Supernova, needed for our present work, are illustrated. 
Both FD and PL energy distributions (labelled $\eta_{FD}$ and $\eta_{PL}$, respectively) 
are shown for three different values of the width parameter $w=0.7$, $w=0.8$ and $w=0.9$
(see Appendix) and for five equivalent parametrizations. From the FD distributions (with 
parameters the temperature T and the width parameter $w$), we see that, as the temperature 
grows the maximum of the distribution shifts to larger $\nu$-energy (at the same time the 
corresponding peak becomes smaller). Also, as the width parameter $w$ grows (keeping the 
same temperature), both the maximum of the distribution shifts to smaller $\varepsilon_{\nu}$ 
and its peak becomes smaller. Furthermore, the degeneracy parameter shifts the spectrum t
o higher energies \cite{tsak_prc2,Jachowicz06}. In this figure, the PL energy distributions 
for the corresponding values of mean neutrino energy $\langle \varepsilon_{\nu}\rangle$, 
are also illustrated ($\langle \varepsilon_{\nu}\rangle$ reflects the depth of the stars 
from which the neutrinos are escaping. We see that, as the $\langle \varepsilon_{\nu}\rangle$ 
grows, the maximum of the distribution shifts to higher $\nu$-energy $\varepsilon_{\nu}$ 
\cite{Raffelt03,Jachowicz06}. 

In Table \ref{Equiv-Param-FD_PL}, the corresponding values of parameters for the equivalent 
FD and PL $\nu$-energy spectra of Fig. \ref{fd-pl}, that have been employed in various SN 
scenarios are shown (for more details see the Appendix and Ref. \cite{tsak_prc2}). It is 
worth mentioning that, due to neutrino oscillations and other phenomena, at any distance 
from the source the SN-$\nu$ spectra can be different compared to those originally produced 
at the core of the collapsing star. It is, however, expected that $\nu$-signals with much 
higher statistics from future galactic SN, may allow us to assess the great number of 
neutrino mixing scenarios. 

It is worth mentioning that, the statistics for the SN 1987A were rather poor, just a few 
dozen $\widetilde{\nu}_e$ events were received within about ten seconds. For the observation 
of the next core-collapse SN-neutrino burst, however, detectors with huge statistics and 
remarkably greater flavour sensitivity are in operation or have been planned to operate in 
the near future \cite{Scholberg-Exp-Statis}. Among those, are the next generation detectors 
HyperKamiokande, Juno, Dune, etc., which aim at measuring, among others, the diffuse SN 
neutrino background \cite{DSNB-Vissani,DSNB-Moller}. 

\section{ Simulated neutrino signals on nuclear detectors }

The features of a neutrino-flux that arrives at a neutrino detector are concealed 
in the nuclear response of the detector-material. In the case of the COBRA detector,
the semi-conductor materials CdTe or CdZnTe contain large portion of $Cd$ isotopes
\cite{Zuber-plb-1,Zuber-plb-2,Zuber-PPNP-10}. 
Our aim in this section is to simulate some of these features by calculating
convoluted cross sections as discussed in Refs. \cite{tsak_prc2,Tsak_AHEP}.      .

The convolution (folding) is carried out with (i) the original 
cross sections obtained in Sect. III, and (ii) the low and intermediate energy 
neutrino spectra of Section IV in order to compute, first, flux averaged total 
cross sections, $\langle \sigma_{tot}\rangle$ and, then, corresponding supernova 
neutrino event rates and fluxes.

\subsection{Flux averaged cross sections for Cd detector materials}

For the coherent channel, which is possible only in neutral current neutrino-nucleus 
reactions studied in this work, the flux averaged cross section $\langle\sigma_{coh}\rangle$ 
is obtained through the folding \cite{Kos-Ose,TSAK-prc1,tsak_prc2}
\begin{equation}
\langle\sigma_{coh}\rangle = \int_{0}^\infty \sigma_{coh}(\varepsilon_{\nu})
\eta (\varepsilon_{\nu}) d\varepsilon_{\nu} \, .
\label{aver-cros-coh}
\end{equation}
For a CdTe or CdZnTe detector material, the flux averaged cross sections, computed 
by inserting in Eq. (\ref{aver-cros-coh}) the $\sigma_{coh}(\varepsilon_{\nu})$ 
from Fig. 2(b) and the $\eta (\varepsilon_{\nu})$ from Figs. 3 and 4, for the 
isotopes $^{112,114,116}$Cd, are listed in Tables \ref{Table2} and \ref{Table3} 
as described below.

In Table \ref{Table2} we list the flux averaged cross sections evaluated by adopting 
the neutrino distributions of the geo-neutrinos (see Fig. \ref{Spectrumms}(a)), the reactor 
neutrinos (see Fig. \ref{Spectrumms}(b)) and the solar neutrinos 
(see Fig. \ref{Spectrumms}(c) for the $^8$B and the $hep$ solar neutrinos).

\begin{table*}
\begin{center}
\begin{tabular}{|ccccccccc|}
\hline\hline
\multicolumn{9}{|c|}{ } \\
\multicolumn{9}{|c|}{ Flux Averaged Cross Sections $\langle \sigma_{coh}\rangle$ $(10^{-39} \, cm^2$)} \\
\hline\hline
\multicolumn{1}{|c|}{  } &\multicolumn{3}{|c|}{  } & \multicolumn{3}{|c|}{ } & \multicolumn{2}{|c|}{  }\\
\multicolumn{1}{|c|}{} &\multicolumn{3}{|c|}{ Geo-Neutrinos } & \multicolumn{3}{|c|}{Reactor
Neutrinos } & \multicolumn{2}{|c|}{ Solar Neutrinos }\\
\hline
\multicolumn{1}{|c|}{  } &\multicolumn{3}{|c|}{  } & \multicolumn{3}{|c|}{ } & \multicolumn{2}{|c|}{  }\\

\multicolumn{1}{|c|}{Isotope} &\multicolumn{1}{|c}{$^{40}$K}  &  \multicolumn{1}{c}{$^{238}$U} & \multicolumn{1}{c|}{$^{232}$Th} & \multicolumn{1}{|c}{$^{235}$U}& \multicolumn{1}{c}{$^{238}$U}& \multicolumn{1}{c|}{$^{239}$Pu}&  \multicolumn{1}{|c}{$^{8}$B} & \multicolumn{1}{c|}{hep} \\
\hline
\hline
\multicolumn{1}{|c|}{  } &\multicolumn{3}{|c|}{  } & \multicolumn{3}{|c|}{ } & \multicolumn{2}{|c|}{  }\\
\multicolumn{1}{|c|}{$^{112}$Cd}& \multicolumn{1}{|c}{0.142}& \multicolumn{1}{c}{1.410}& \multicolumn{1}{c|}{0.911}& \multicolumn{1}{|c}{0.180}& \multicolumn{1}{c}{0.477}& \multicolumn{1}{c|}{9.001}& \multicolumn{1}{|c}{7.970}& \multicolumn{1}{c|}{9.333} \\
\hline
\multicolumn{1}{|c|}{  } &\multicolumn{3}{|c|}{  } & \multicolumn{3}{|c|}{ } & \multicolumn{2}{|c|}{  }\\
\multicolumn{1}{|c|}{$^{114}$Cd} &\multicolumn{1}{|c}{0.151} & \multicolumn{1}{c}{1.504} & \multicolumn{1}{c|}{0.973} & \multicolumn{1}{|c}{0.192} & \multicolumn{1}{c}{0.509} & \multicolumn{1}{c|}{9.604} & \multicolumn{1}{|c}{8.504} & 
\multicolumn{1}{c|}{9.957} \\
\hline
\multicolumn{1}{|c|}{  } &\multicolumn{3}{|c|}{  } & \multicolumn{3}{|c|}{ } & \multicolumn{2}{|c|}{  }\\
\multicolumn{1}{|c|}{$^{116}$Cd} &\multicolumn{1}{|c}{0.161} &  \multicolumn{1}{c}{1.602} & \multicolumn{1}{c|}{1.036} & \multicolumn{1}{|c}{0.205}& \multicolumn{1}{c}{0.542}& \multicolumn{1}{c|}{1.023}& \multicolumn{1}{|c}{9.055} & 
\multicolumn{1}{c|}{10.598} \\
\hline
\hline
\end{tabular}
\end{center}
\caption{ Calculated values for the flux averaged coherent cross sections $\langle \sigma_{coh}\rangle$ 
(in units $10^{-39}$ cm$^2)$ for $^{112}$Cd, $^{114}$Cd and $^{116}$Cd isotopes. The neutrino sources
distributions of neutrino beams coming from: (i) Geo-neutrinos, (ii) Reactor neutrinos and (iii) Solar 
neutrinos have been used in the folding procedure.
\label{Table2}}
\end{table*}

In the last two columns of Table \ref{Table3} we tabulate the $\langle\sigma_{coh}\rangle$ 
calculated for the distributions of Eqs. (\ref{Lab-dist-nnu}) and (\ref{Lab-dist-nmu2}), 
i.e. the $\nu$-spectra produced by pion/muon decay at rest (DAR). In the first 
three columns of this Table, the flux averaged cross sections refer to various 
supernova neutrino scenarios described by the equivalent FD and PL distributions of 
Fig. \ref{fd-pl}. The corresponding parameters are listed in Table \ref{Equiv-Param-FD_PL}. 

In supernova neutrino scenarios, 
usually average $\nu$-energies between 10 $\le \langle\varepsilon_\nu\rangle\le$ 12 MeV 
are employed for the description of $\nu_e$ neutrinos, average energies between
$15\le\langle\varepsilon_\nu\rangle\le 18$ MeV for $\widetilde{\nu}_e$ anti-neutrinos, and 
average energies between 22 $\le\langle\varepsilon_\nu\rangle\le$ 26 MeV for $\nu_x$ and
$\widetilde{\nu}_x$, with x = $\mu$, $\tau$ \cite{Janka89,Raffelt03,Raffelt-proc03,Jachowicz06}. 

\begin{table*}
\begin{center}
\begin{tabular}{|ccccccccccccc|}
 \hline\hline
&&&&&&&&&&&&\\
\multicolumn{13}{|c|}{ Flux Averaged Cross Sections $\langle \sigma_{coh}\rangle$ $(10^{-39} \, cm^2$)} \\
\hline\hline
\multicolumn{1}{|c||}{ } &\multicolumn{8}{c}{ } & \multicolumn{4}{|c|}{} \\
\multicolumn{1}{|c||}{ Isotope } &\multicolumn{8}{c}{ Supernova Neutrinos } & \multicolumn{4}{|c|}{Pion-muon DAR
Neutrinos} \\
\hline
\multicolumn{1}{|c||}{} & \multicolumn{4}{c}{} & \multicolumn{4}{|c}{} & \multicolumn{2}{|c}{} &\multicolumn{1}{c}{} &
\multicolumn{1}{c|}{} \\
\multicolumn{1}{|c||}{\textbf{}} &\multicolumn{4}{c}{Fermi-Dirac (FD)} & \multicolumn{4}{|c}{Power Law (PL)} &
\multicolumn{2}{|c}{} &\multicolumn{1}{c}{${\nu_e}$} Spect. & \multicolumn{1}{c|}{${\widetilde{\nu}_\mu}$ Spect. }  \\
\hline
\multicolumn{1}{|c||}{} &\multicolumn{4}{c}{} & \multicolumn{4}{|c}{} & \multicolumn{2}{|c}{} &\multicolumn{1}{c}{} & \multicolumn{1}{c|}{} \\
\multicolumn{1}{|c||}{} &\multicolumn{1}{c}{ $T = $ } &3.10& 4.14&6.20 & \multicolumn{1}{|c}{ $\langle\varepsilon_{\nu} \rangle
= $ } & 12 & 16 & 24 &\multicolumn{2}{|c}{}&\multicolumn{1}{c}{}& \multicolumn{1}{c|}{} \\
\hline
\multicolumn{1}{|c||}{} &\multicolumn{4}{c}{} & \multicolumn{4}{|c}{} & \multicolumn{2}{|c}{} &\multicolumn{1}{c}{} & \multicolumn{1}{c|}{} \\

\multicolumn{1}{|c||}{\textbf{$^{112}$Cd}} & & 2.484 & 4.184 & 8.132 & \multicolumn{1}{|c}{} & 2.489 & 4.180 & 8.142 & \multicolumn{1}{|c}{}& &12.338 & 14.960 \\
\hline

\multicolumn{1}{|c||}{}  &\multicolumn{4}{c}{}  &  \multicolumn{4}{|c}{} & \multicolumn{2}{|c}{} &\multicolumn{1}{c}{} & \multicolumn{1}{c|}{} \\

\multicolumn{1}{|c||}{\textbf{$^{114}$Cd}} & &2.648 & 4.458 & 8.648 & \multicolumn{1}{|c}{} & 2.653 & 4.453 & 8.658 & \multicolumn{1}{|c}{}& &13.110 & 15.881 \\

\multicolumn{1}{|c||}{} &\multicolumn{4}{c}{} & \multicolumn{4}{|c}{} & \multicolumn{2}{|c}{} &\multicolumn{1}{c}{} & \multicolumn{1}{c|}{} \\

\multicolumn{1}{|c||}{\textbf{$^{116}$Cd}} & & 2.817 & 4.739 & 9.178 & \multicolumn{1}{|c}{} & 2.823 & 4.734 & 9.189 & \multicolumn{1}{|c}{}& &13.801& 16.824 \\
\hline
\hline
\end{tabular}
\end{center}
\caption{ Flux averaged coherent cross sections $\langle \sigma_{coh}\rangle$, as in Table \ref{Table2} 
but now referred to: (i) three different Supernova neutrino spectra determined from the parameters of: 
(a) Fermi Dirac parametrizations and (b) Power-Law parametrizations, 
and (ii) the energy spectra of Pion/muon decay at rest (DAR) neutrinos.
\label{Table3} }
\end{table*}

Due to the dominance of the coherent channel throughout the region of the incoming neutrino 
energy $\varepsilon_{\nu}$ of our present calculations, the flux averaged coherent cross 
section $\langle\sigma_{coh}\rangle$ may be even two or three orders of magnitude 
larger than the total incoherent cross section $\langle\sigma_{tot}^{incoh}\rangle$ 
\cite{TSAK-prc1,tsak_prc2,prc3}.

\subsection{ Number of events in $\nu$-detectors}\label{Number_of_events}
 
The present theoretical results may be connected with current neutrino experiments 
relying on Cd isotopes as detection materials, and specifically the COBRA experiment
at Gran Sasso \cite{Zuber-plb-1,Zuber-plb-2,Zuber-PPNP-10}, as follows. By 
using the flux averaged cross sections $\langle\sigma(\varepsilon_\nu)\rangle$ of 
Table \ref{Table3}, for instance those referred to the SN neutrinos of
the $^{112,114,116}$Cd isotopes, we estimate (potentially
detectable) neutrino fluxes $\Phi_\nu$ that should arrive at each detector to create 
some typical scattering event rates $N_{ev}$ in the COBRA detector. 

In general, the event rate $N_{ev}$ is related to the flux $\Phi_\nu$ reaching the 
nuclear detector with the expression \cite{TSAK-prc1,Bal-Ydr-Kos-11,Tsak_AHEP}
\begin{equation}
\frac{dN_\nu}{dt}\equiv N_{ev}=N_{Cd}\sigma_{tot}(\varepsilon_\nu)\Phi_\nu(\varepsilon_\nu) \, . 
\label{Nevents}
\end{equation}
We note that, experimentalists use the definition 
$$
N_{ev} = \epsilon N_{Cd}\sigma_{tot}(\varepsilon_\nu)\Phi_\nu(\varepsilon_\nu) \, ,
$$ 
which takes into account the detection efficiency $\epsilon$ (usually equal to 
$\epsilon \approx$ 80 - 90\%) of the specific detector.
Here, we assume a COBRA detector of mass $m_{det}$=100 kg and two cases of detector materials,
i.e the semiconductors (a) CdZnTe and (b) CdTe \cite{Zuber-plb-1,Zuber-plb-2,Zuber-PPNP-10}. 

In the first step, we choose three SN neutrino scenarios in which the mean energies are:
(i) $\langle \varepsilon_\nu \rangle = 12$ MeV (corresponding to SN electron neutrinos ${\nu}_e$), 
(ii) $\langle \varepsilon_\nu \rangle = 16$ MeV (corresponding to SN electron anti-neutrinos
$\widetilde{\nu}_e$), and 
(iii) $\langle \varepsilon_\nu \rangle = 24$ MeV (corresponding to SN $\nu_x$, $\widetilde{\nu}_x$, 
with $x=\mu, \tau$ (anti)neutrinos of heavy leptons).

Then, based on Eq. (\ref{Nevents}), we perform calculations assuming a total mass 
100 kg of CdZnTe as COBRA detector which translates, for example, to approximately $m_{Cd}=10.6$ 
kg mass of $^{114}$Cd isotope or equivalently a number of $^{114}$Cd atoms (nuclei) equal 
to $N_{Cd} \equiv N_{^{114}Cd}= 94.17 N_{Avog}$. 

In Eq. (\ref{Nevents}), as total neutrino scattering cross sections, $\sigma_{tot}(\varepsilon_\nu)$ 
we employ the values of flux averaged cross sections $\langle \sigma_{coh}\rangle$ of Table \ref{Table3}
obtained through PL distribution for SN neutrino spectra (they refer to the three mean energies chosen 
above).

Finally, we choose four typical detection rates $N_{ev}$ as:
(a) $N_{ev}$=1 event s$^{-1}$=3.15$\times 10^{7}$ events y$^{-1}$, 
(b) $N_{ev}$=1 event hr$^{-1}$=8.76$\times 10^{3}$ events y$^{-1}$, 
(c) $N_{ev}$=1 event d$^{-1}$=3.65$\times 10^{2}$ events y$^{-1}$, 
and (d) $N_{ev}$= 12 events y$^{-1}$
and (from Eq. (\ref{Nevents}) we compute the corresponding SN $\nu$ fluxes $\Phi_{\nu}$.

In a similar way, assuming that the COBRA detector contains 100 kg of the material CdTe, 
we find 13.5 kg $^{114}$Cd or about $N_{Cd}\equiv N_{^{114}Cd} = 120.11 N_{Avog}$ atoms (nuclei)
are contained in the second semiconductor material of COBRA detector. By performing similar 
calculations for the same SN scenarios and the same, as before, set of detection rates 
$N_{ev}$, we find the corresponding fluxes $\Phi_\nu$ reaching the COBRA CdTe detector.

By performing the steps we followed for $^{114}$Cd, for the other two Cd-isotopes, 
$^{112}$Cd and $^{116}$Cd, the resulting neutrino fluxes, for the chosen SN neutrino 
scenarios are listed in Table \ref{Table4} (last four columns). Such results are useful 
for future use of the Cd materials in astrophysical neutrino detection. It should be 
stressed that, next generation experiments may be effective in the detection of much 
weaker signals (higher sensitivity, larger detector mass, etc.).

The above neutrino fluxes are of the same order with those of the Spallation Neutron 
Source (SNS) at ORLaND, Oak Ridge \cite{ORLaND,Avignone03,Burman-03}. We mention that 
the COHERENT experiment at Oak Ridge, with a 14.57 kg of CsI scintillator detector, by 
using an SNS $\nu_\mu$ neutrino flux (coming from $\pi$-decay at rest) as high as 
$\Phi_\nu^{COH} = 1.7 \times 10^{11} \nu_{\mu}/{\textrm cm}^2$ s, has measured 142 CEvNS 
events within a period of 308.1 live days (at a distance of L = 19.3 m from the source) 
\cite{CEnN_neut}. These results translate to event rate $N_{ev}^{COH}$ = 168/y $\nu_{\mu}$ 
neutrinos.

From the results of Table \ref{Table4}, we may define the ratio $N_{ev}/\Phi_\nu$ for the 
COHERENT experiment ($R^{COH}$) and for a special $\nu_\mu$ neutrino case of the COBRA
experiment ($R^{COB}$). For a comparison of these two experiments, we choose, for example, 
the results referred to the $^{112}$Cd isotope of CdTe material of the COBRA detector (sixth
line from the beginning of Table \ref{Table4} refers to $\nu_{\mu}$ neutrinos). 
From these two ratios we find that $R = R^{COH}/R^{COB} = 98.95/2.94 \approx 34$, which means 
that, for the chosen SN neutrino scenario, the COBRA detector may observe 12 $\nu_{\mu}/y$ only 
if its mass is equal to m $\sim$ 34 times larger than the assumed above 100 kg, i.e. only if the
COBRA detector has a huge total mass $m_{det} = 3.4$ t CdTe material (we mention that, in the 
assumed scenario, the SN $\nu_{\mu}$ neutrinos correspond a Temperature T=24 MeV, see one before 
last column of Table \ref{Table4}). This example indicates also the corresponding cost for
detector improvement so as to be able to record neutrino signals coming from interesting
astrophysical sources. 

We should finally note that, in this work the detection efficiency $\epsilon$ has not 
been considered (equivalently we assumed $\epsilon=1$). Also, the neutrino mixing has 
not been accounted for which means that we assumed the neutrino spectra arrived at the 
nuclear detector are described by PL distributions (as in stars interior) of the same 
values of the parameters.

\begin{table*}
\begin{center}
\begin{tabular}{|l|c|c|c|c|r|r|r|r|r|}
\hline
\hline
\multicolumn{10}{|c|}{   }  \\
\multicolumn{10}{|c|}{Supernova neutrino coherent fluxes $\Phi_\nu$ (s$^{-1} \, {\textrm cm}^{-2})$ 
and event rates $N_{ev}$}\\
\hline
 \multicolumn{5}{|c|} { } & & & & & \\
 \multicolumn{5}{|c|} { } & 
$N_{ev}$ = 1/s & $N_{ev}$ = 1/hr & $N_{ev}$ = 1/d & $N_{ev}$ = 12/y & $N_{ev}^{COH}$ = 168/y \\
\hline
& & &                           &   &         &     &         &       &   \\
Isotope & Detector & Atoms ($N_{Avog}$) & m (kg) & $\langle \varepsilon_\nu \rangle$ (MeV) &
$\Phi_\nu(\times 10^5) $ &$\Phi_\nu(\times 10^9)$&$\Phi_\nu(\times 10^{10} )$&$\Phi_\nu(\times 10^{11} )$ 
&$\Phi_\nu^{COH}(\times 10^{11} )$\\
\hline
& & &                             & 12 &  5.14   & 1.85 &  4.44  & 13.33 &  \\
$^{112}$Cd &  CdTe &100.08 &11.25 & 16 &  3.06   & 1.10 &  2.65  &  7.94 &  \\
& & &                             & 24 &  1.57   & 0.57 &  1.36  &  4.08 & 1.70 \\
\hline
& & &                             & 12 &  5.15   & 1.85 &  4.45  & 13.34 &  \\
$^{112}$Cd & CdZnTe &100.05&11.25 & 16 &  3.06   & 1.10 &  2.64  &  7.94 &  \\
& & &                             & 24 &  1.57   & 0.57 &  1.36  &  4.08 & 1.70 \\
\hline
\hline
& & &                             & 12 &  3.98   & 1.45 &  3.44  & 10.32 &  \\
$^{114}$Cd &  CdTe  &121.29&13.50 & 16 &  2.37   & 0.86 &  2.05  &  6.15 &  \\
& & &                             & 24 &  1.22   & 0.45 &  1.05  &  3.16 & 1.70 \\
\hline
& & &                             & 12 &  4.26   & 1.85 &  3.67  & 11.00 &  \\
$^{114}$Cd & CdZnTe &121.26&10.60 & 16 &  2.53   & 1.10 &  2.18  &  6.55 &  \\
& & &                             & 24 & 12.98   & 0.57 &  1.12  &  3.36 & 1.70 \\
\hline
\hline
& & &                             & 12 & 14.10   & 5.08 & 12.19  & 36.56 &  \\
$^{116}$Cd &  CdTe  & 32.18& 3.62 & 16 &  8.41   & 3.03 &  7.27  & 21.80 &  \\
& & &                             & 24 &  4.33   & 1.56 &  3.74  & 11.23 & 1.70 \\
\hline
& & &                             & 12 & 16.00   & 5.76 & 13.83  & 41.48 &  \\
$^{116}$Cd & CdZnTe & 32.18& 3.62 & 16 &  9.53   & 3.43 &  8.23  & 24.70 &  \\
& & &                             & 24 &  4.89   & 1.76 &  4.23  & 12.66 & 1.70 \\
\hline 
\hline
\end{tabular}
\caption{Neutrino fluxes $\Phi_\nu(\varepsilon_\nu)$ and corresponding event rates $N_{ev}$ 
estimated to be recorded on $^{112,114,116}$Cd isotopes of two detector materials (CdTe and 
CdZnTe) of the COBRA experiment \cite{Zuber-plb-1,Zuber-plb-2,Zuber-PPNP-10}. They refer to 
the case of supernova neutrinos with mean energies $\langle \varepsilon_\nu \rangle$ 
= 12, 16 and 24 MeV. $N_{Avog}$ is the Avogadro's number. In the last column, $N_{ev}^{COH}$ 
and $\Phi_\nu^{COH}$ describe COHERENT experiment values (see the text).} 
\label{Table4}
\end{center}
\end{table*}

\section{Summary and Conclusions}

In this work, we present original neutrino-nucleus cross sections obtained with realistic 
nuclear structure calculations (use of the QRPA method) for scattering of low and intermediate 
energy neutrinos off the $^{112,114,116}$Cd isotopes. These Cd-isotopes are contents 
(with large abundance) of the detector materials of the COBRA detector at Gran Sasso. 
The neutrino energy assumed covers currently interesting laboratory (reactor, pion/muon 
decay at rest neutrinos) and Astrophysical (solar, supernova and Earth) neutrino sources. 
Laboratory neutrino beams are important tools for studying standard and non-standard 
neutrino physics while astrophysical neutrinos are key particles in investigating the 
structure and evolution of stars as well to deepen our knowledge on the fundamental 
neutrino-nucleus interactions.

By utilizing the convolution procedure, we calculated flux averaged cross sections and event 
rates for the above $\nu$-sources based on specific spectral distributions describing supernova 
neutrino energy spectra, solar neutrinos, geo-neutrinos and laboratory neutrinos as well as 
reactor neutrinos and pion-muon-stopped neutrinos. The flux-averaged total coherent cross sections, 
$\langle \sigma_{coh}\rangle$, reflect the mean neutrino signals generated in several terrestrial 
detectors ($^{112,114,116}$Cd) from such $\nu$-sources. Important connection of our present results 
with current experiments may also be achieved through the evaluation of the neutrino scattering 
event rates on Cd detectors.

The estimated neutrino fluxes and scattering event rates for Cd-isotopes, contents of the CdTe 
and CdZnTe materials of the COBRA detector at LNGS, may support this experiment to reach its goal 
in searching for neutrino observation and detection of rare events (double beta decay, etc). 

\section{Acknowledgements}  

The present research was financially supported (V. Tsakstara and J. Sinatkas) by the Department 
of Informatics Engineering of the Technological Institute of Western Macedonia. Also, Dr. Odysseas 
Kosmas wishes to acknowledge the support of EPSRC via grand EP/N026136/1 "Geometric Mechanics of 
Solids". 


\vspace*{0.40 cm}


{\large \bf Appendix}

\vspace*{-0.3 cm}


\subsection{Three-point formulas for empirical energy gaps $\Delta^{exp}_{n,p}$ of neutrons and protons }

The empirical energy gaps for neutrons, $\Delta^{exp}_n$, and protons, $\Delta^{exp}_n$, needed at
the BCS level to construct the ground state wave function of the detector nucleus $(A, Z)$, are computed 
through the respective separation energies for neutrons, $S_n$ or protons, $S_p$ of the isotope 
$(A, Z)$ and also those of the neighbouring nuclear isotopes with $N \pm 1$ neutrons or $Z \pm 1$ 
protons, respectively, 
by employing the expressions
\begin{eqnarray}
\Delta^{exp}_n = - \frac{1}{4} \left[ S_n (N - 1, Z) - 2 S_n (N, Z) + S_n (N + 1, Z) \right] \, \\
\Delta^{exp}_p = - \frac{1}{4} \left[ S_p (N, Z - 1) - 2 S_p (N, Z) + S_p (N, Z + 1) \right] \, 
\end{eqnarray}
The above equations are known as the three-point formulas (see, e.g. Ref. \cite{Bohr-Motel,prc3}). 

\subsection{ Normalization of the $\nu$ energy distributions $\eta (\varepsilon_\nu)$ adopted in this work }

The distributions $\eta (\varepsilon_\nu)$ adopted in the present work (see Sect. IV), are considered 
to be normalized in such a way that
\begin{equation}
\int_{0}^\infty \eta (\varepsilon_\nu) d\varepsilon_\nu = 1 .
\end{equation}

For example, in the case of $\eta_{\nu_e} (\varepsilon_{\nu})$ of Fig. \ref{Spectrumms}(d), the 
normalization gives
$$ 
(96/M_\mu^4) \left[ M_\mu \int_0^{M_\mu/2}\varepsilon_\nu^2 d \varepsilon_\nu - 
2 \int_0^{M_\mu/2} \varepsilon_\nu^3 d \varepsilon_\nu \right]  = 1
$$
where we have used $\varepsilon_\nu^{min} = 0$ and $\varepsilon_\nu^{max} = {M_\mu/2}$

\subsection{ Parametrization of Supernova neutrino energy spectra }

The Fermi-Dirac (FD) and Power-law (PL) energy distribution are commonly used in
Supernova neutrino parametrizations. Both the FD and PL yield very similar 
distributions characterized by the temperature $T$ or the average energy
$\langle\varepsilon_\nu\rangle$ and the width $w$ of the spectrum is defined as
$$
w = \sqrt{ \langle\varepsilon_\nu^2\rangle - \langle\varepsilon_\nu\rangle^2 }/w_0
$$
where $w_0 = \varepsilon_\nu/\sqrt{3}$ is the width of the identical FD and PL
distributions \cite{tsak_prc2}.

\subsubsection{Fermi-Dirac (FD) energy distribution }

By introducing the degeneracy parameter $n_{dg}$ (equal to the ratio of the chemical potential
$\mu$ divided by the neutrino temperature $T$, i.e. $n_{dg} = \mu/T$), the Fermi-Dirac energy 
distribution reads
\begin{equation}
\eta_{FD}[x,T,n_{dg}] = F(n_{dg}) \,
 \frac{1}{T}\frac{x^2}{1+e^{{(x - n_{dg})}}} \, , \qquad  x = \frac{\varepsilon_\nu}{T} \, .
 \label{2-param-FD}
\end{equation}
In this case, the width of the spectrum is reduced compared to the corresponding
thermal spectrum (pinching effect). The normalization constant  $F_2(n_{dg})$ of
this distribution depends on the degeneracy parameter $n_{dg}$ and is given by 
the relation
\begin{equation}
\frac{1}{F(n_{dg})} \equiv \int_0^\infty
\frac{x^{2}}{e^{x-n_{dg}}+1} dx \, . \label{norm-coef}
\end{equation}
Inserting Eq. (\ref{norm-coef}) into Eq. (\ref{2-param-FD}), we take
\begin{equation}
\eta_{FD}[\varepsilon_{\nu},T,n_{dg}]=\left[\int_0^\infty
\frac{x^{2}}{e^{x-n_{dg}}+1} dx \right]^{-1}\,
\frac{(\varepsilon_{\nu}^2/T^3)}{1+e^{(\varepsilon_{\nu}/T-n_{dg})}}
\, . \label{mm}
\end{equation}

\subsubsection{Power-law energy distribution }


The SN-neutrino energy spectra can be fitted by using a Power-Law energy distribution 
of the form \cite{Raffelt03}
\begin{equation}
\eta_{PL}[\langle \varepsilon_{\nu}\rangle ,\alpha] = C \left(
\frac{\varepsilon_{\nu}}{\langle \varepsilon_{\nu} \rangle}
\right)^\alpha e^{-(\alpha+1)(\varepsilon_{\nu}/\langle
\varepsilon_{\nu}\rangle)} \, , \label{2-param-PLL}\
\end{equation}
where $\langle \varepsilon_{\nu} \rangle $ is the neutrino mean energy. The 
parameter $\alpha$ adjusts the width of the spectrum (see text). The normalization 
factor $C$, is calculated from the normalization condition
\begin{equation}
C \int_0^\infty \left(
\frac{\varepsilon_{\nu}}{\langle \varepsilon_{\nu} \rangle}
\right)^\alpha e^{-(\alpha+1)(\varepsilon_{\nu}/\langle
\varepsilon_{\nu} \rangle)} d \varepsilon_{\nu} \,  =1 \, .
\label{PL-norm}\
\end{equation}
From the later equation we find
\begin{equation}
C = \frac{(\alpha+1)^{\alpha+1}}{\Gamma(\alpha+1) \langle
\varepsilon_{\nu} \rangle} \, ,
\end{equation}
therefore, Eq. (\ref{2-param-PLL}) becomes
\begin{equation}
\eta_{PL}[\langle \varepsilon_{\nu} \rangle ,\alpha] =
\frac{(\alpha+1)^{\alpha+1}}{\Gamma(\alpha+1)} \,
\frac{\varepsilon_{\nu} ^\alpha}{\langle \varepsilon_{\nu}\rangle^{\alpha+1}}
e^{-(\alpha+1)(\varepsilon_{\nu} /\langle \varepsilon_{\nu}
\rangle)} \, . \label{Nornal-PL-distr}\
\end{equation}





\begin{thebibliography}{10}
\bibitem{Ejiri-PR00} H. Ejiri, Nuclear spin isospin responces for low-energy neutrinos, Phys. Rep. (2000) \textbf{338}:265--351.
\bibitem{Ejiri-plb-02} H. Ejiri, J. Engel, N. Kudomi, Supernova-neutrino studies with $^{100}$Mo, Phys. Lett. B (2002) \textbf{530}:27--32.
\bibitem{Zuber-plb-1} K. Zuber, COBRA-double beta decay searches using CdTe detectors, Phys. Lett. B (2001) \textbf{519}:1--7.
\bibitem{Zuber-plb-2} K. Zuber, Spectroscopy of low energy solar neutrinos using CdTe detectors, Phys. Lett. B (2003) \textbf{571}:148--154.

\bibitem{DonPe} T.W. Donnelly and R.D. Peccei, Neutral current effects in nuclei, Phys. Rep. (1979) \textbf{50}:185 .
\bibitem{Kos-Ose} T.S. Kosmas and E. Oset, Charged current neutrino-nucleus reaction cross sections at intermediate energies, Phys. Rev. C (1996) \textbf{53}:1409--1415.
\bibitem{kolbe-kos} E. Kolbe and T.S. Kosmas, Recent highlights on neutrino-nucleus interactions, Springer Trac. Mod. Phys. (2000) \textbf{163}199--225.

\bibitem{CEnN_neut} D. Akimov {\it et al.}, Observation of coherent elastic neutrino-nucleus scattering, Science (2017) \textbf{357}:1123--1126.
\bibitem{Scholberg} K. Scholberg, Prospects for measuring coherent neutrino-nucleus elastic scattering at a stopped-pion neutrino source, Phys. Rev. D (2006) \textbf{73}:033005-1--9.
\bibitem{Scholb_IoP-2010} K. Scholberg, Coherent elastic neutrino-nucleus scattering, J. Phys. Conf. Ser. (2015) \textbf{606}:012010-1--10.

\bibitem{TSAK-prc1} V. Tsakstara and T.S. Kosmas, Low-energy neutral-current neutrino scattering on $^{128,130}$Te isotopes, Phys. Rev. C (2011) \textbf{83}:054612-1--13.
\bibitem{kos-eric-11} V. Tsakstara, T.S. Kosmas, J. Wambach, Studying low-energy astrophysical neutrinos with neutrino nucleus cross-section calculations and beta beam neutrino
spectra, Prog. Part. Nucl. Phys. (2011) \textbf{66}:424--429.
\bibitem{tsak_prc2} V. Tsakstara and T.S. Kosmas, Analyzing astrophysical neutrinos through realistic nuclear structure calculations and the convolution procedure, Phys. Rev. C (2011) \textbf{84}:064620-1--14.

\bibitem{KamLAND} S. Abe {\it et al.}, Precision Measurement of Neutrino Oscillation Parameters with KamLAND (KamLAND Collaboration), Phys. Rev. Lett. (2008) \textbf{100}:221803-1--5.
\bibitem{KamLAND-11} A. Gando {\it et al.}, Partial radiogenic heat model for Earth revealed by geoneutrino measurements (KamLAND Collaboration), Nature Geo. (2011) \textbf{4}:647--651.
\bibitem{Borexino-Col} G. Bellini {\it et al.}, Observation of Geo-Neutrinos (Borexino Collaboration), Phys. Lett. B (2010) \textbf{687}:299--304.
\bibitem{Borexino} G. Bellini {\it et al.}, Precision Measurement of the $^{7}$Be Solar Neutrino Interaction Rate in Borexino (Borexino Collaboration), Phys. Rev. Lett. (2011) \textbf{107}:141302-1--5.
\bibitem{Borexino-13} G. Bellini {\it et al.}, Measurement of geo-neutrinos from $1353$ days of Borexino (Borexino Collaboration), Phys. Lett. B (2013) \textbf{722}:295--300.
\bibitem{Zub-SNO+} K. Zuber, Status of the double beta experiment COBRA, Prog. Part. Nucl. Phys. (2006) \textbf{57}:235--240.
\bibitem{LENA} M. Wurm {\it et al.}, The next-generation liquid-scintillator neutrino observatory LENA, Astropart. Phys. (2012) \textbf{35}:685--732.

\bibitem{Kolbe-Lag-Pin-03} E. Kolbe, K. Langanke, G. Martinez-Pinedo, and P. Vogel, Neutrino-nucleus reactions and nuclear structure, J. Phys. G (2003) \textbf{29}:2569--2596.
\bibitem{Athar-Sing} M.Sajjad Athar and S.K. Singh, $\nu_{e}(\bar{\nu}_{e})-^{40}$Ar absorption cross sections for supernova neutrinos, Phys. Lett. B (2004) \textbf{591}:69--75.

\bibitem{bet-beam-Zuc} P. Zucchelli, A novel concept for a $\bar{\nu}_{e}/\nu_{e}$ neutrino factory: the beta-beam, Phys. Lett. B (2002) \textbf{532}:166--172.
\bibitem{bet-beam-Vol} C. Volpe, What about a beta beam facility for low-energy neutrinos?, J. Phys. G. (2004) \textbf{30}:L1--L6 .
\bibitem{Volpe07} C. Volpe, Topical Review on Beta-beams, J. Phys. G (2007) \textbf{34}:R1--R44.
\bibitem{Zuber-PPNP-10} K. Zuber, The status of the COBRA double-beta-decay experiment, Prog. Part. Nucl. Phys. (2010) \textbf{64}:267--269. 
 
\bibitem{Smpon-Ody-2015} T. Smponias, O.T. Kosmas, High Energy Neutrino Emission from Astrophysical Jets in the Galaxy, Advances in High Energy Physics (2015) Article ID 921757. 
\bibitem{Smpon_Ody-2017} T. Smponias, O.T. Kosmas, Neutrino Emission from Magnetized Microquasar Jets, Advances in High Energy Physics (2017) Article ID 4962741.
\bibitem{Smpon-Ody-2018} O.T. Kosmas, T. Smponias, Simulations of Gamma-Ray Emission from Magnetized Microquasar Jets (2018) Article ID 9602960.

\bibitem{IceCube_2015} M.~G. Aartsen {\it et al.}, Search for Prompt Neutrino Emission from Gamma-Ray Bursts with IceCube (IceCube Collaboration), Astrophys. J. Lett. (2015) \textbf{805}:L5.
\bibitem{KM3Net_2016} S. Adrian-Martinez {\it et al.}, Letter of intent for KM3NeT 2.0 (KM3Net Collaboration), J. Phys. G: Nucl. Part. Phys. (2016) \textbf{43}:084001-(130pp).

\bibitem{Tsak_AHEP} V. Tsakstara, Convoluted-Signals on $^{114}$Cd Isotope from Astrophysical and Laboratory Neutrino Sources, Advances in High Energy Physics (2015) Article ID 632131. 
\bibitem{Kosmas-94} T.S. Kosmas, J.D. Vergados, O. Civitarese, and Amand Faessler, Study of the muon number violating ($\mu^{-}$, $e^{-}$ conversion in a nucleus by using quasi-particle RPA, Nucl. Phys. A (1994) \textbf{570}:637--656.
\bibitem{Kosmas01}T.S. Kosmas, Exotic $\mu^{-}\rightarrow e^{-}$ conversion in nuclei: energy moments of the transition strength and average energy of the outgoing $e^{-}$, Nucl. Phys. A (2001) \textbf{683}:443--462.

\bibitem{ody05} I.G. Tsoulos, O.T. Kosmas, and V.N. Stavrou, DiracSolver: A tool for solving the Dirac equation, Computer Physics Communications (2019) \textbf{236}:237--243.
\bibitem{ody01} O. Kosmas and S. Leyendecker, Analysis of higher order phase fitted variational integrators, Advances in Computational Mathematics (2016) \textbf{42}:605--619.
\bibitem{ody02} O. Kosmas and S. Leyendecker, Variational integrators for orbital problems using frequency estimation, Advances in Computational Mathematics (2019) \textbf{45}:1--219.
\bibitem{ody03} O. Kosmas and D. Papadopoulos, Multisymplectic structure of numerical methods derived using non standard finite difference schemes, Journal of Physics: Conference Series \textbf{490}:012205.
\bibitem{ody04} O.T. Kosmas, Charged Particle in an Electromagnetic Field Using Variational Integrators, AIP Conference Proceedings (2011) \textbf{1389}:1927.
\bibitem{Vogel-Beacom} P. Vogel and J.F. Beacom, Angular distribution of neutron inverse beta decay $\bar{\nu}_{e}+\vec{p}e^{+}+n$, Phys. Rev. D (1999) \textbf{60}:053003-1--10.
\bibitem{Dye-2011} S. Dye, Geoneutrinos and the radioactive power of the Earth, Rev. Geophys. (2012) \textbf{50}:3007.
\bibitem{Fiorentini-2003} G. Fiorentini, E. Mantovani, and B. Ricci, Neutrinos and energetics of the Earth, Phys. Lett. B (2003) \textbf{557}:139--146.
\bibitem{Fiorentini-2010} G. Fiorentini {\it et al.}, Nuclear physics for geo-neutrino studies, Phys. Rev. C (2010) \textbf{81}:034602-1--9.

\bibitem{ORLaND} F.T. Avignone and Y.V. Efremenko, ORLaND - a neutrino facility at the Spallation Neutron Source, Nucl. Phys. B (Proc. Supl.) (2000) \textbf{87}:304--308 .
\bibitem{Avignone03} F.T. Avignone and Y.V. Efremenko, Neutrino-nucleus cross-section measurements at intense, pulsed spallation sources, J. Phys. G (2003) \textbf{29}:2615--2628.
\bibitem{Burman-03} R.L. Burman and W.C. Louis, Neutrino physics at meson factories and spallation neutron sources, J. Phys. G (2003) \textbf{29}:2499--2512.

\bibitem{Papoul-Kosm} D.K. Papoulias and T. S. Kosmas, Nuclear aspects of neutral current non-standard $\nu$-nucleus reactions and the role of the exotic $\mu^{-}\rightarrow e^{-}$ transitions experimental limits, Phys. Lett. B (2014) \textbf{728}:482--488.
\bibitem{Papoul_tsk_AHEP-2015} D.K. Papoulias and T.S. Kosmas, Standard and Nonstandard Neutrino-Nucleus Reactions Cross Sections and Event Rates to Neutrino Detection Experiments, Advances in High Energy Physics (2015) Article ID 763648.


\bibitem{Engel-91} J. Engel, Nuclear form-factors for the scattering of weakly interacting massive particles, Phys. Lett. B (1991) \textbf{264}:114--119.
\bibitem{De-Vries} H.De Vries, C.W. De Jager, and C. De Vries, Nuclear charge-density-distribution parameters from elastic electron scattering, Atomic Data and Nuclear data tables (1987) \textbf{36}:495--536.
\bibitem{Drukier} A. Drukier and L. Stodolsky, Principles and applications of a neutral-current detector for neutrino physics and astronomy, Phys. Rev. D (1984) \textbf{30}:2295-2309.

\bibitem{Horowitz-PRC-2012} C J. Horowitz {\it et al.}, Weak charge form factor and radius of $^{208}$Pb through parity violation in electron scattering, Phys. Rev. C (2012) \textbf{85}:032501-1--5.
\bibitem{tsk-NPA-92} T.S. Kosmas and J.D. Vergados, Nuclear densities with fractional occupation probabilities of the states, Nucl. Phys. A (1992) \textbf{536}:72--86.
\bibitem{Chiang-Oset-Kosmas} H.C. Chiang {\it et al.}, Coherent and incoherent $(mu-, e-)$ conversion in nuclei, Nucl. Phys. A (1993) \textbf{559}:526--542. 

\bibitem{Bachall} J.N. Bahcall, Neutrino Astrophysics, Cambridge University Press, 1989 (Reprint 1990), Melbourne, Australia.

\bibitem{Omar-1} J. Barranco, O.G. Miranda, T.I. Rashba, Probing new physics with coherent neutrino scattering off nuclei, Journal of High Energy Physics (2005) \textbf{05}:021.  
\bibitem{Omar-2} J. Barranco, A. Bolanos, O.G. Miranda, and T.I. Rashba, Tensorial NSI and Unparticle physics in neutrino scattering, Int. J. of Mod. Phys. A (2012) \textbf{27}:1250147.

\bibitem{JDV-Avig} J.D. Vergados, F.T. Avignone, and I. Giomataris, Coherent neutral current neutrino-nucleus scattering at a spallation source: A valuable experimental probe, Phys. Rev. D (2009) \textbf{79}:113001-1--8.
\bibitem{Giom-JDV} Y. Giomataris and J.D. Vergados, A network of neutral current spherical TPCs for dedicated supernova detection, Phys. Lett. B (2006) \textbf{634}:23--29.
\bibitem{JDV-Giomat} J.D. Vergados and Y. Giomataris, Dedicated supernova detection by a network of neutral current spherical TPC detectors, Phys. Atom. Nucl. (2007) \textbf{70}:140--149.

\bibitem{Vogel-Engel} P. Vogel and J. Engel, Neutrino electromagnetic form factors, Phys. Rev. D (1989) \textbf{39}:3378--3383.
\bibitem{Kolbe-96} E. Kolbe, Differential cross sections for neutrino scattering on $^{12}$C, Phys. Rev. C (1996) \textbf{54}:1741--1748.

\bibitem{Lang-Pin-03} K. Langanke and G. Martinez-Pinedo, Nuclear weak-interaction processes in stars, Rev. Mod. Phys. (2003) \textbf{75}:819--862.
\bibitem{Juoda-Lang} A. Juodagalvis, K. Langanke, G. Martinez-Pinedo, W.R. Hix, D.J. Dean, and J.M. Sampaio, Neutral-current neutrino-nucleus cross sections for $A \sim 50-65$ nuclei, Nucl. Phys. A (2005) \textbf{747}:87--108.
\bibitem{Lang-08} K. Langanke, Weak interaction, nuclear physics and supernovae, Acta Phys. Polon. B (2008) \textbf{39}:265--282.
\bibitem{Hax-87} W.C. Haxton, Nuclear response of water Cherenkov detectors to supernova and solar neutrinos, Phys. Rev. D (1987) \textbf{36}:2283--2292.

\bibitem{Don-Wal72} T.W. Donnelly and J.D. Walecka, Semi-leptonic weak and electromagnetic interactions in nuclei with application to $^{16}$O, Phys. Lett. B (1972) \textbf{41}:275--280.
\bibitem{Don-Wal73} T.W. Donnelly and J.D. Walecka, Elastic magnetic electron scattering and nuclear moments, Nucl. Phys. A (1973) \textbf{201}:81--106.

\bibitem{prc3} V. Tsakstara and T.S. Kosmas, Nuclear responses of $^{64,66}$Zn isotopes to supernova neutrinos, Phys. Rev. C (2012) {\bf86}:044618-1--10.
\bibitem{Bal-Ydr-Kos-11} K.G. Balasi, E. Ydrefors, and T. S. Kosmas, Theoretical study of neutrino scattering off the stable even Mo isotopes at low and intermediate energies, Nucl. Phys. A (2011) \textbf{868-869}:82--98.

\bibitem{Anderson-2012} A.J. Anderson {\it et al.}, Measuring active-to-sterile neutrino oscillations with neutral current coherent neutrino-nucleus scattering, Phys. Rev. D (2012) \textbf{86}:013004-1--11.
\bibitem{Gouvea-neutrinos} A. de Gouvea {\it et al}, Working Group Report: Neutrinos (Intensity Frontier Neutrino Working Group), arXiv:1310.4340 [hep-ex] (2013).       
\bibitem{Bachall-Holstein} J.N. Bahcall and B.R. Holstein, Solar neutrinos from the decay of 8B, Phys. Rev. C (1986) \textbf{33}:2121--2127. 
\bibitem{Bachall-Urlich} J.N. Bahcall and R.K. Urlich, Solar models, neutrino experiments, and helioseismology, Rev. Mod. Phys. (1988) \textbf{60} :297--372.
\bibitem{Bachall-05a} J.N. Bahcall {\it et al.}, Helioseismological Implications of Recent Solar Abundance Determinations, Astrophys. J. (2005) \textbf{618}:1049--1056.

\bibitem{BOONE} W.C. Louis, Searches for muon-to-electron (anti) neutrino flavor change, Prog. Part. Nucl. Phys. (2009) \textbf{ 63}:51--73.

\bibitem{Davis-Vogel-79} B.R. Davis, P. Vogel, F.M. Mann, and R.E. Schenter, Reactor antineutrino spectra and their application to antineutrino-induced reactions, Phys. Rev. C (1979) \textbf{19}:2259--2266.
\bibitem{Bugey_experiment} Y. Declais {\it et al.}, Study of reactor antineutrino interaction with proton at Bugey nuclear power plant, Phys. Lett. B (1994) \textbf{338}:383--389.
\bibitem{Tengblad-89} O. Tengblad {\it et al.}, Integral $\bar{\nu}$-spectra derived from experimental $\beta$-spectra of individual fission products, Nucl. Phys. A (1989) \textbf{503}:136--160.

\bibitem{TEXONO-Kerman} S. Kerman {\it et al.}, Coherency in neutrino-nucleus elastic scattering (TEXONO Collaboration), Phys. Rev. D (2016) \textbf{93}:113006.
\bibitem{TEXONO-Sevda} B. Sevda {\it et al.}, Constraints on Scalar-Pseudoscalar and Tensorial Non-Standard Interaction and Tensorial Unparticle Couplings from Neutrino-Electron Scattering (TEXONO Collaboration), Phys. Rev. D (2017) \textbf{95}:033008.
\bibitem{MINER-Agnolet} G. Agnolet {\it et al.}, Background Studies for the MINER Coherent Neutrino Scattering Reactor Experiment (MINER Collaboration), Nucl. Instrum. Methods Phys. Res. A (2017) \textbf{853}:53--60.

\bibitem{Bethe90} H.A. Bethe, Supernova mechanisms, Rev. Mod. Phys. (1990) \textbf{62}:801--866.
\bibitem{Lank-prl} C. Fr\"{o}hlich {\it et al.}, Neutrino-induced nucleosynthesis of $A>64$ nuclei: the vp process, Phys. Rev. Lett. (2006) \textbf{96}:142502.
\bibitem{Janka95} H.-Th. Janka, and B. Mueller, Neutrino-driven type-II supernova explosions and the role of convection, Phys. Rep. (1995) \textbf{256}:135--156.
\bibitem{Janka07} H.-Th. Janka, K. Langanke, A. Mareka, G. Martinez-Pinedo, and B. Mullera, Theory of core-collapse supernovae, Phys. Rept. (2007) \textbf{442}:38--74.
\bibitem{Janka89} H.T. Janka and W. Hillebrand, Neutrino emission from type II supernovae -- an analysis of the spectra, Astron. Astrophys. (1989) \textbf{224}:49--56.
\bibitem{Raffelt03} M.T. Keil, G.G. Raffelt, and H.-T. Janka, Monte Carlo study of supernova neutrino spectra formation, Astrophys. J. (2003) \textbf{590}:971--991.
\bibitem{Raffelt-proc03} G.G. Raffelt, M.T. Keil, R. Buras, H.-T. Janka, and M. Rampp, Supernova neutrinos: flavor-dependent fluxes and spectra, Proc. Neutrino Oscillations and Their Origin (2004) \textbf{03}:380--387.

\bibitem{Duan08} H. Duan {\it et al.}, Stepwise spectral swapping with three neutrino flavors, Phys. Rev. D (2008) \textbf{77}:085016.   
\bibitem{Jachowicz06} N. Jachowicz and G.C. McLaughlin, Reconstructing Supernova-Neutrino Spectra using Low-Energy Beta Beams, Phys. Rev. Lett. (2006) \textbf{96}:172301-1--4.
\bibitem{Vtsak-Zn-fold} V. Tsakstara, T.S. Kosmas, P.C. Divari and J. Sinatkas,  The interpretation of SN?v signals in terrestrial experiments through the folding procedure, AIP Conf. Proc. (2009) \textbf{1180}:140--144.
\bibitem{TSAK10-erice2-2010} V. Tsakstara and T.S. Kosmas, Neutrino-nucleus reactions in terrestrial experiments and astrophysics, Prog. Part. Nucl. Phys. (2010) \textbf{64}:407--410.

\bibitem{Scholberg-Exp-Statis} K. Scholberg,  Supernova signatures of neutrino mass ordering, J.Phys. G (2018) \textbf{45}:014002.
\bibitem{DSNB-Vissani} F. Vissani and G. Pagliaroli, The diffuse supernova neutrino background: expectations and uncertainties derived from SN1987A, Astronomy \& Astrophysics (2011) \textbf{528}:L1. 
\bibitem{DSNB-Moller} K. Moeller, A.M. Suliga, I. Tamborra, P.B. Denton, Measuring the supernova unknowns at the next-generation neutrino telescopes through the diffuse neutrino background, Journal of Cosmology and Astroparticle Physics (2018) \textbf{1805}:066.

\bibitem{Bohr-Motel} A. Bohr and B.R. Mottelson, Nuclear Structure, Benjamin, New York, 1969.
\end{thebibliography}
\end{document}